# Elusive Longer-Run Impacts of Head Start: Replications Within and Across Cohorts

By Remy J.-C. Pages, Dylan J. Lukes, Drew H. Bailey and Greg J. Duncan*

January 31, 2020

WORKING PAPER.

PLEASE DO NOT CITE OR CIRCULATE WITHOUT PERMISSION.

*Using an additional decade of CNLSY data, this study replicated and extended Deming's (2009) evaluation of Head Start's life-cycle skill formation impacts in three ways. Extending the measurement interval for Deming's adulthood outcomes, we found no statistically significant impacts on earnings and mixed evidence of impacts on other adult outcomes. Applying Deming's sibling comparison framework to more recent birth cohorts born to CNLSY mothers revealed mostly negative Head Start impacts. Combining all cohorts shows generally null impacts on school-age and early adulthood outcomes.*

* Remy J.-C. Pages and Dylan J. Lukes are co-equal first authors. Before collaborating, both individuals were separately and contemporaneously working on individual papers which substantially overlapped. After careful consideration all parties agreed to collaborate and produce a single unified paper. We thank Felipe Barrera-Osorio, Damon Clark, David Deming, Chloe Gibbs, Josh Goodman, Eric Taylor for feedback on prior drafts and presentations of this project. Contact information for each of these authors can be viewed below: School of Education, University of California, Irvine, 3200 Education, Irvine, CA 92697 (e-mail: rpages@uci.edu) and School of Education, Harvard University, 13 Appian Way, Cambridge, MA 02138 (email: dylanlukes@g.harvard.edu).

## Introduction

Understanding the causal effects of early childhood programs implemented at scale on long-term adult outcomes is challenging. However, since early childhood is considered by many economists to be a key launching period for life-long human capital accumulation (e.g., Cunha et al., 2006; Chetty et al., 2011; Currie & Almond, 2011; Heckman & Mosso, 2014; Hoynes, Schanzenbach & Almond, 2016), considerable attention has been devoted to research attempting to estimate the short and longer-run impacts of early education programs (Duncan & Magnuson, 2013).

One such early childhood education program is Head Start, the U.S.'s oldest and largest early childhood education program to be offered at scale. Given this, it is hardly surprisingly a great deal of the research has been devoted to it.[1] Studies of the longer-run impacts of Head Start attendance have shown generally positive, although sometimes mixed, results (Garces, Thomas & Currie (GTC), 2002; Ludwig & Miller, 2007; Deming, 2009; Carneiro & Ginja, 2014; Bauer & Schanzenbach, 2016; Thompson, 2018). In a recent study analyzing data from the Panel Study of Income Dynamics survey, Johnson and Jackson (2019) suggested that some of these inconsistencies can be attributed to complementarities between Head Start attendance and subsequent K-12 spending.[2]

One important study on the long-term impacts of Head Start attendance is Deming (2009). Using data from the National Longitudinal Survey of Youth 1979 Children and Young Adults (CNSLY), his paper builds on the approach of Currie and Thomas (1995; also, GTC, 2002) by comparing both school-age and young adulthood outcomes between children who attended Head Start and their siblings who either attended other non-Head Start preschools or did not attend any preschool education. Most of the cohorts analyzed in Deming (2009) were born between 1976 and 1986 and had outcomes tracked through the survey's 2004 interviewing wave.[3] The study found that, compared with siblings who did not attend any preschool, children who attended Head Start averaged 8.5 percentage points (*pp*) higher rates of high school graduation and 0.23 standard deviation (*SD*) higher scores on an index of adult outcomes.[4] Deming (2009) is noteworthy both for its sibling comparison design, which controls for some unmeasured time-invariant factors of the family environment, and because of its use of a reasonably large and relatively recent national longitudinal sample followed through childhood into early adulthood.

However, recent research calls into question the sibling comparison design both in terms of its external validity (Miller, Shenhav & Grosz (MSG), 2019) and potential to produce biased estimates from sibling spillover effects (Heckman & Karapakula, 2019). To correct for non-random selection into the family fixed effects (FFE) identifying sample, MSG (2019) have found that reweighting on observables attenuates many of the original Deming (2009) FFE estimates of Head Start's impact on long-term outcomes. MSG (2019) document similar attenuated FFE estimates of Head Start's long-term impact in the Panel Study of Income Dynamics (PSID). Highlighting threats to construct validity, Heckman & Karapakula (2019) found siblings who participated in the Perry Preschool Project had large positive spillovers on their non-participating siblings. This was particularly true for male siblings.

Importantly, evidence of Head Start's lasting positive impact into adulthood is not limited to Deming (2009) or the FFE design. Over the past several decades, a sizable body of evidence that leverages a variety of empirical methods including FFE, regression discontinuity (RD) and difference-in-difference (DID) has accumulated indicating Head Start's ability to improve adolescent and longer-term outcomes (Currie & Thomas, 1995, 1999; GTC, 2002; Ludwig &

Miller, 2007; Carneiro & Ginja, 2014; Bauer & Schanzenbach, 2016; Bailey, Sun & Timpe, 2018; Thompson, 2018; Barr & Gibbs, 2019; MSG, 2019). Most consistent across this body of research was Head Start's positive impact on educational attainment, health outcomes and reduced criminal activity with estimated impacts tending to be larger and more robust for males, siblings from earlier birth cohorts and those born to mothers with less than a high school education.

On whole, these studies predominantly focused on 1970s and 1980s birth cohorts. Notable exceptions include Carneiro and Ginja's (2014) RD analysis of 1977-1996 birth cohorts and Bauer and Schanzenbach's (2016) FFE analysis of 1970-1990 birth cohorts. Although not part of their main results, Barr and Gibbs' (2019) supplementary FFE analysis sample (contained in their appendix) included CNLSY 1970-1992 birth cohorts. Results were mixed on Head Start's impact for more recent birth cohorts. Carneiro and Ginja (2014) indicated that a robustness check showed that the positive effects for males age 12-13 in their overall sample were driven by the earlier 1980s birth cohorts. Similarly, in an appendix FFE analysis, Barr and Gibbs (2019) found no significant impact of Head Start on high school graduation, some college, crime, teen parenthood or their index of adulthood outcomes. However, Head Start impacts were positive and significant for males on high school graduation, crime and their index of adulthood outcomes (Barr & Gibbs, 2019). In both cases, each overall sample included birth cohorts from 1970s through the early to mid-1990s. In contrast, Bauer and Schanzenbach's (2016) FFE analysis found positive impacts of Head Start on high school graduation, some higher education, post-secondary completion, self-control index, self-esteem index and positive parenting index. These results more closely followed Deming (2009) and included birth cohorts up to 1990.[5] A detailed synthesis of these studies and more—including birth cohorts analyzed; identification strategy; and findings—can be found in the Appendix.

## Present study

The present work builds upon this rich existing literature by expanding Deming's (2009) evaluation of Head Start's longer-run impacts. By appending 10 additional years to the original 1976-1986 birth cohorts analyzed in Deming (2009), we were able to estimate impacts on outcomes measured later in adulthood and not previously considered: educational attainment, college graduation, and earnings. Second, the additional data provided us with an opportunity to apply the methods used in Deming (2009) to 10 additional birth cohorts in the NLSY to address whether his results generalized to cohorts born to older mothers and into somewhat different historical conditions. Third, we estimated impacts on both school-age and adulthood outcomes for a sample combining all possible cohorts to provide estimates based on the broadest population base.

We found that extending the measurement period for Deming's cohorts and early-adult outcomes decreased the estimated impact on the adulthood summary index (ASI) of Head Start attendance relative to not attending any preschool program from 0.23 *SD* to 0.17 *SD* (standard error (*SE*) = 0.07). Of the longer-run outcomes we were able to consider, the largest impact of attending Head Start was on years of completed schooling (0.30 years; *SE* = 0.15). This is notable, and taken by itself, could indicate a sizable return on investment for the program. However, we estimated relatively small, nonsignificant impacts on gains on other later life outcomes including college graduation and earnings.[6] For the cohorts born after Deming's cohorts, Head Start impacts were mostly null and sometimes negative. In fact, positive impacts on ASI generated by Deming's cohorts were matched by nearly symmetric negative impacts for the complement cohorts (-0.15 *SD; SE* = 0.07). For the final sample that combined the two sets of cohorts, the point estimate of Head Start's impact on the ASI was close to zero and not statistically significant.

In light of recent work by MSG (2019), following these initial analyses we checked whether our FFE identifying samples exhibited "selection into identification" (SI) across a variety of observable characteristics including family size and mother's age at child's birth. Finding evidence of SI in both the Deming and combined cohorts, we used the one-step reweighting-on-observables procedure outlined in MSG (2019) to correct for any potential bias. Similar to MSG (2019), after reweighting the Deming cohort we found attenuated estimates of Head Start's impact on long-term outcomes. However, for the combined cohorts sample we found limited evidence that reweighting attenuated Head Start impact estimates on long-term outcomes.

Our paper concludes with a discussion of what is driving these cross-cohort differences in Head Start's impacts. While we found differences in baseline human capital between cohorts, we found no evidence that the impact of Head Start varied for different levels of human capital within cohorts. Similarly, we observed differences in other pre-treatment covariates between cohorts, but also found limited evidence that they drove variation in Head Start impacts. Finally, to better understand whether the effects of Head Start were changing across cohorts, we performed a Blinder-Oaxaca decomposition (Blinder, 1973; Oaxaca, 1973) using a threefold decomposition (Jann, 2008). In line with the above analyses, the results from this exercise also highlighted key differences between Deming's and complement cohort samples, but importantly showed how these differences – most notably in the pre-treatment index and mother's age at child's birth – were associated with variation in estimated Head Start impacts across cohorts.

These results indicated that if mother's age at child's birth was set fixed at Deming's cohort level, ASI mean for complement cohort's Head Start attendees would be similar to that of Deming's cohort counterparts.

Thus, although the past several years has seen a resurgence of research on the long-run impacts of Head Start, this study adds value in several notable ways. First, we used a well-established FFE design, which hitherto has estimated positive long-run outcomes of Head Start, to estimate predominantly negative or null long-run outcomes of the program and showed how, if at all, these results were sensitive to "selection into identification" by performing the one-step reweighting-on-observables procedure as outlined in MSG (2019). Second, despite major changes to the program and social context of Head Start-eligible children during this period, ours is the first paper to have estimated the impact of Head Start for the most recent set of CNLSY birth cohorts and to compare their program effects with those of earlier cohorts.[7] Finally, we attempted to reconcile why Head Start impacts were different across cohorts, finding suggestive evidence that between cohort differences in the ages of mothers at the time of their child's birth played an important explanatory role.

## Method

**Head Start Program Background**

Part of the Johnson administration's Great Society policies, Head Start was launched in 1965 to provide educational and health-related services to children living in low-income families. As of 2017, about 900,000 children were enrolled in Head Start, 97 percent of whom were between the ages of 3 and 5, at an annual cost of around $9 billion in federal funding (U.S. Department of Health and Human Services (DHHS), 2018). Enrollment and funding have varied greatly since Head Start's 1965 inception. Participation grew until the early 1980s, plateaued

through the early 1990s and then grew again when funded enrollments almost doubled (i.e., from around 500,000 in 1990 to 900,000 in 2000). Appropriations (in 2018 USD) grew from about $3 billion in 1990 to $9 billion in 2000. After 2000, both enrollment and inflation-adjusted funding remained steady (DHHS, 2018).

Between the 1989-1990 (which are typical Head Start attendance years for Deming's cohorts) and 1996-1997 (which are typical attendance years for our complement cohorts), enrollment increased by about 60 percent. However, the proportion of teachers or assistant teachers with at least a Child Development Associate credential increased very little – by about 5 percentage points (*pp*) over this period (DHHS, 2018). More generally, the 1990s and 2000s were a time of rapid increases in preschool enrollment, including Head Start, but also state-run pre-kindergarten programs (Duncan & Magnuson, 2013).

**Data**

Figure 1 provides an overview of birth years and years in which childhood and adult outcomes are measured for the two sets of cohorts that form our analytic samples. Deming's cohorts were born between 1970 and 1986 and attended Head Start no later than 1990. Moreover, Deming's sibling fixed effects analyses were estimated for a sample of siblings discordant on Head Start attendance and who enrolled in Head Start no later than 1990. Deming's sample eligibility rules were: 1) at least two children aged 4 or older by 1990 within the same family; and 2) at least one pair of siblings in a family had to be discordant across Head Start, other preschool, or neither statuses. The median age of individuals in Deming's analytic sample was 23 years (21 and 25 years for first and third quartile, respectively) by 2004, the most recent CNLSY survey round year available for his study.

[Insert Figure 1 here]

For our complement and combined cohorts, Deming's sample restrictions were moved forward by ten years: samples were restricted to siblings who were at least 4 by 2000 (i.e., at least 19 by 2014). Sample restrictions produced sample sizes of N = 1,251 for Deming's cohorts, N = 2,144 for our complement cohorts and N = 3,768 for our combined cohorts.[8] It is important to note that the sampling design of the CNLSY (i.e., all children were born to women who were between ages 14 and 22 in 1979) led children in our complement cohort to be born to older mothers than is the case for children in Deming's cohort. Later, consider the role this factor may have played in explaining differences in Head's Start impact between the Deming and complement cohorts. Further, because we wanted to both estimate Head Start impacts on educational attainment, college graduation and earnings, and assess the impacts' robustness on ASI, we both replicated and extended Deming's analysis of this cohorts up to 2014, the latest CNLSY survey round year available to us at the time of our analyses.

**Family background statistics**

In Table 1, household characteristics are presented by cohort and preschool status (Head Start vs. the counterfactual of no preschool); permanent income; maternal education and cognitive test score; and grandmothers' highest grade completed.[9] Across these variables and for all three cohorts, there was a clear pattern of selection of more disadvantaged children into Head Start for samples of siblings under rule 1 only—a less restricted sample, more representative of the CNLSY sample—and samples with rule 2 added (i.e., the fixed effects subsamples). Discrepancies between the two samples were small, suggesting that the demographic characteristics of the fixed effect subsamples were similar to the less restricted, larger samples.[10]

[Insert Table 1 here]

As shown in the column 'Difference HS-None'—reporting mean differences in standard deviation units for Deming's cohorts, the complement cohorts and the combined cohorts, respectively—selection into Head Start was similarly associated with socioeconomic disadvantage for Deming's cohort as well as for the complement cohort. For example, Head Start participants had a 0.44 *SD* lower permanent income and a 0.59 *SD* lower maternal AFQT than children not attending any preschool. Overall, then, Head Start children came from relatively more disadvantaged households.[11] As noted by Deming (2009), because his cohort of Head Start participants had been born to younger mothers (their median age was 20), they might have benefited more from the program (which, in addition to early education, includes services for parents). In contrast, for the complement cohort, mothers were older (median age was 28), and household characteristics more favorable on all of the dimensions included in Table 1.

**Outcomes**

As part of our replication and extension of Deming (2009), we assessed the impact of Head Start on the same set of three test scores, two nontest outcomes and six young adulthood outcomes for each of the Deming, complement and combined cohorts. The three test scores covered ages 5-14 and included: the Peabody Picture Vocabulary Test (PPVT), the Peabody Individual Achievement Math (PIATMT) subtest, and the PIAT Reading Recognition (PIATRR) subtest. Following Deming (2009), due to the biannual survey design of the CNLYS we pooled PPVT tests scores of five and six-year-olds to get the first post-Head Start score for each child in our sample. Both the PIATMT and PIATRR were administered annually for respondents ages 5-14 and resulted in considerably more observations compared to the PPVT.

The two nontest outcomes covered ages 7-14 and included: grade retention and learning disability diagnosis. As in Deming (2009), grade retention is a dichotomous variable based on survey respondents' answers to whether their child has ever been retained at grade-level while in school. This question was asked biannually in the NLSY from 1988 to 2014. Grade retention was coded as a one if parents ever answered "yes" to this question across any of the survey years. Learning disability was based off a "yes" or "no" NLSY survey question that asked parents if their child had a learning disability. We coded our learning disability variable as one if respondents ever answered "yes" to this question, discounting a small number of children that were diagnosed with a learning disability prior to age 5.

Finally, as in Deming (2009) we included the same six young adulthood outcome variables: high school graduation, teen parenthood, some college attended, idleness, involvement with the justice system, and poor health status. All outcomes were measured up to the CNLSY 2014 survey-round. Individuals were considered "idle" if they were not enrolled in school or had reported zero annual earnings – by 2004 for the Deming cohort and by 2014 for the complement cohort. The "involvement with the justice system" variable was constructed as a dichotomous variable, coded as one if a respondent ever answered "yes" to any survey question related to conviction, probation, sentencing and prison. Teen parenthood was operationalized also with an indicator equal to one if a respondent's age at the birth of their first child was before 20-years-old and applied to both female and male respondents. Finally, our poor health status variable was constructed by averaging a respondent's self-reported health framed by a 1-5 Likert scale (lower responses equating to poorer self-reported health). Poor health status was flagged by a dichotomous variable coded as one, if respondents average self-reported health score was less than three on the Likert scale.

Just as in Deming (2009), to reduce the risk of multiple-inference inflated Type I errors and mitigate measurement error we constructed summary indices for the three test scores, an index for the two nontest score outcomes, and a final adulthood summary index (ASI) for the six young adulthood outcomes. Outcomes were normalized to have mean of zero and standard deviation of one, with positive index values signal "good" outcomes and negative index values signal "bad" outcomes. The final index was then created by taking a simple average of all the normalized and, where appropriate, re-signed outcomes.

Comparing the distribution of these outcomes across the Deming and complement cohorts, we found evidence of substantial distributional shifts often favoring the complement cohort (Table S5 & Figure S6). These between-cohort distributional changes might explain why Head Start impacts were dramatically different for the more recent set of siblings in the complement cohort. To illustrate, on the ASI, complement cohort siblings attending Head Start and not attending preschool were respectively 0.22 $SD$ and 0.50 $SD$ higher than the Deming cohort siblings. Similarly, although there were a small and marginally significant 0.10 $SD$ difference on the cognitive test index across cohorts for Head Start attendees, that difference was more pronounced for the No Preschool status children (0.30 $SD$; $p <.001$). Given these distributional shifts in outcomes were between cohorts across time and not within a cohort across Head Start treatment statuses, these results were not inconsistent with later findings that the Deming and complement cohorts showed no signs of within-family selection into Head Start or No Preschool status (Table S7).

Finally, the longer time series of NLSY data enabled us (but not Deming) to estimate Head Start effects for Deming's cohort on completed years of schooling and college graduation, and on earnings.[12] The earnings composite for each sample member was obtained by first pooling all person-year earnings observations (in 2014 USD) and then regressing them on dummy-variable indicators for birth cohort and calendar year to purge earnings of birth cohort and measurement year effects.[13] From the coefficients in this regression, we generated a set of person-year earnings residuals for all individuals in the analysis sample. We then averaged these earnings residuals for each individual, added them to the grand mean earnings in the sample, and took the natural logarithm of this earnings average.

**Empirical Strategy**

As noted, families selecting into Head Start were relatively more disadvantaged on a series of selected household characteristics. Consequently, Head Start estimates relative to other preschool status based on cross-family variation may be negatively biased. A family fixed-effects design mitigates some of these biases by separating the potentially confounding influence of family environment variance shared among siblings from estimations of interest. This was the empirical strategy undertaken in Deming (2009), which we reproduced in the present study and formalized in the same fashion:

(1) $$Y_{ij} = \alpha + \beta_1 HS_{ij} + \beta_2 PRE_{ij} + \delta \mathbf{X}_{ij} + \gamma_j + \varepsilon_i.$$

In this model, $i$ and $j$ respectively index individuals and families. Thus, $HS_{ij}$ ($PRE_{ij}$) stands as an indicator for participation in Head Start (Preschool) where $\beta_1$ ($\beta_2$) denotes Head Start (Preschool) impact estimates on outcome $Y_{ij}$, for some sibling $i$ within family $j$, relative to a sibling (within family $j$) attending neither. Next, $\mathbf{X}_{ij}$ represents the vector of 'pre-treatment' family covariates pertaining to sibling $i$ within family $j$; family $j$ fixed-effect is captured by $\gamma_j$, while $\varepsilon_i$ represents sibling $i'$s residual.

**Selection Bias**. Within-family comparisons remove the effects of time-invariant family characteristics on siblings' outcomes. There remains, however, a strong possibility of within-family selection bias. Reasons for different within-family patterns of care in early childhood were not recorded in the CNLSY. To mitigate such potential for bias, Deming (2009) opted for a series of sibling-specific family-level covariates—the ones represented by the vector **X** in equation (1)—measured before or at Head Start preschool program eligibility age (3 years old). We examined these covariates for the complement and combined cohorts and tested whether siblings within a given family systematically differed on these covariates.[14]

Within our family fixed-effects framework, each covariate was thus regressed on siblings' preschool status, either 'Head Start' or 'Preschool'. A statistically significant and substantial variation from 'No preschool' (the reference status) would then signal a potential selection bias regarding the relation between that pre-treatment characteristic and the regressed on preschool status. These estimates are reported in the online appendix, Table S7.

Focusing first on the complement cohort, siblings attending Head Start were on average older by one year, and by almost two years for siblings attending other preschools, than their counterparts not attending any preschool program. This was consistent with the probability being greater for first-born sibling to be enrolled in preschool, by 10 *pp* for Head Start enrollees, and by 28 *pp* for other preschool participants. Both groups were 6 to 8 *pp* less likely to receive maternal care from birth to age three, and so more likely to receive care from a non-relative (5 and 6 *pp*, respectively). Attrition was low for both the complement and the combined cohorts, averaging about 4 and 3 percent, respectively. Moreover, both Head Start and other preschool participants were somewhat less likely (by about 3 and 2 *pp*, respectively) to be part of observations lost to attrition.

To characterize selection bias as it pertains to overall disadvantage, a summary index of all pre-treatment covariates was constructed in the same way as the multiple outcomes adulthood index described earlier.[15] For all cohorts, as with the covariates, the pre-treatment index was regressed on the two preschool indicators, keeping the no preschool status as the reference category. We found that Head Start and other preschool within-family effects on the pre-treatment index were close to zero and never statistically significant. As with Deming's (2009) selection bias analysis (which we replicated, see online appendix Tables S8-S9), we could not reject the null hypothesis of equality between preschool statuses.

While there was no evidence of parental selection into Head Start (or any other preschool status) within each cohort and household, Figure 2 illustrates what might be one explanation for cross-cohort differences of Head Start impacts on longer-run adulthood outcomes. Complement cohort respective kernel densities of pre-treatment index scores were shifted to the right, i.e., towards more favorable household characteristics for the complement cohort. Complement cohort siblings having attended Head Start later would have then, on average, benefited from more household resources. Compared with Deming's cohort, such a shift might stand as a potential substitute for whatever impact the program would have otherwise yielded.

[Insert Figure 2 here]

## Results

Each of the following subsections is organized by cohorts. We first present Head Start impacts on the adulthood summary index (ASI), along with its individual composing outcomes. Second, longer-run Head Start impacts on ASI, educational attainment, college graduation and earnings are described. Third, estimates for school age outcomes are shown. Robustness checks and a reconciliation of results are presented in the final subsection.

**Head Start Impacts on ASI**

In Table 2, the family fixed effects model was implemented in steps—with three model specifications; repeated across the complement and the combined cohorts—to gauge the relative directions of biases from observed covariates and unobserved household-level confounders.[16] Model (1) included no family fixed effects but included the pre-treatment covariates (Table S7), along with household predictors (Table 1)—namely, standardized permanent income; maternal AFQT score; one indicator for maternal high school graduation and one for some college attendance. By contrast, model (2) includes only family fixed effects. Model (3) includes both fixed effects and pre-treatment covariates. Moving from model (1) to model (2), the explained variance ($R^2$) was larger for all cohorts. Hence, error variance from unobserved variables was smaller than that from the selected observed variables. Moreover, within the $R^2$ column, including pre-treatment covariates to the fixed effects model—i.e., moving from model (2) to model (3)—added some precision to the estimates (explained variance increasing from 0.64 to 0.69; from 0.71 to 0.73; and from 0.61 to 0.63 for Deming's, complement and combined cohorts, respectively).[17]

[Insert Table 2 here]

From the middle panel, for the complement cohort, Head Start impacts on ASI were negative at -0.15 *SD* (significant at the 5 percent level). This value was in clear contrast ($p < .01$) with Deming's cohort estimate of 0.17 *SD* ($SE = 0.07$). In the bottom panel, for the combined cohorts model (3), no Head Start impacts' estimates were statistically significant; most were negative and close to zero.[18]

We investigated the decrease in Head Start impact on ASI from 0.23 *SD*—in Deming (2009); outcomes measured up to CNLSY 2004 survey-round—to 0.17 *SD*; outcomes measured up to 2014). The change was due in part to impacts on the indicator 'Idle' changing sign and ceasing to be statistically significant: by 2004, Deming's cohort Head Start participants were 7 *pp* (*SE* = 0.04) less likely to be 'idle'; this impact had disappeared by 2014 (-3 *pp*; *SE* = 0.04). Thus, with the passage of the additional decade, Head Start participants were not, on average, better positioned to pursue a college degree or to have a job, relative to their siblings not having attended any preschool program.

Figure 3 shows Head Start impacts on all individual outcomes composing ASI.[19] For Deming's cohort, 'Poor health status' stayed favorable by 5 *pp* (*SE* = 0.03), decreasing slightly from 2004 (7 *pp*; *SE* = 0.03). Impacts on 'Some college attended' rose to statistical significance (11 *pp*; *SE* = 0.04).[20] On 'Crime', Head Start participants did not appear to have had more involvement with the justice system than their siblings. Yet, impacts on teenage parenthood shifted unfavorably. Since impacts on ASI are based on scores averaged across the composing indicators, Deming's cohort overall decline on this index were captured by changes on the individual outcomes just described.

[Insert Figure 3 here]

In contrast to Deming's cohort, Head Start impacts on the complement cohort were mostly negative and larger in absolute value; when positive, they were smaller in absolute value. Head Start's estimated impact on 'Idle' was relatively large, negative (-0.08; *SE* = 0.03) and significant at the 1 percent level. Thus, in the complement cohort, siblings who have attended Head Start were less likely by about 8 *pp* to be employed or enrolled in school (by age 19 or older), compared with their siblings who received home care. Impact on 'Some college attended' went in the opposite (negative) direction for the complement cohort (-0.07 *SD*; *SE* = 0.04), as well as for 'Crime' (reversed scaled, -0.03; *SE* = 0.03).[21] In sum, the discrepancies between the two cohorts over Head Start impacts on these individual outcomes are aligned with the difference observed earlier over ASI (Table 2). Once more, impact estimates for the combined cohorts sample were small and never statistically significant.

Finally, consistent with Anderson's (2008) study of early childhood interventions life cycle impacts, females appeared to have benefited more than males from Head Start, across the board of outcomes considered here (see online appendix, Tables S10-S13). Over the extended ASI (Deming's cohort), Head Start impact was estimated at 0.23 *SD* (*SE* = 0.11) for females versus 0.10 *SD* (*SE* = 0.10) for males.[22] Further, females possibly carried most of Head Start impact on educational attainment with an estimate at 0.34 *SD* (*SE* = 0.21) against 0.27 *SD* (*SE* = 0.21) for males (online appendix, Table S13).

**Head Start Impacts on Longer-Run Outcomes**

Head Start longer-run impacts are displayed in Figure 4 (the complete set of estimates is presented in online appendix Table S13). As described previously, impacts on ASI declined from Deming's published results (2009) as his study's cohort grew older by a decade (second bar from the top in Figure 4). Yet, by 2014, Head Start attendees went to school 0.3 years longer than their siblings not attending any preschool. This positive, potentially important impact, however, did not translate into either higher college graduation rate or to significantly higher adulthood earnings.[23]

[Insert Figure 4 here]

**Head Start Impacts on School Age Outcomes**

Could Head Start impacts on school age outcomes explain the cohort differences in adult outcomes shown above? For example, are Head Start impacts on achievement generally positive for Deming's cohorts but negative for the complement cohorts? Although a full econometric mediation analysis (e.g., Heckman & Pinto, 2015) was not the focus of this article, school age outcomes might nonetheless be considered as potential mediators (e.g., as cognitive or noncognitive inputs) impacting adulthood outcomes. Estimating Head Start impacts on these earlier outcomes could thus be informative about the processes underlying the pattern of later impacts.[24]

As shown in Figure 5, estimates from Deming (2009) and our replication of Deming (2009) were aligned. For the complement cohort, patterns of impacts on school age outcomes mirrored impacts on the adulthood outcomes: they went in the opposite direction. This was also the case for the nontests index (-0.15 *SD*; *SE* = 0.08), with Head Start's impact on the learning disability diagnosis indicator (reverse scaled; -0.04 *SD*) being statistically significant at the 5 percent level. This is in line with the adverse impacts recorded on the Behavioral problem index (reverse scaled; -0.07 *SD*; *SE* = 0.05). Yet, we could not detect any Head Start impact on 'Grade retention' (whereas for Deming's cohort, the impact of being grade retained was at a lesser 7 *pp*, significant at the 10 percent level).[25] Regarding the cognitive tests index, the relatively sustained gains generated by Head Start for Deming's cohort (0.11 *SD*; *SE*= 0.06) did not reflect those for the complement cohort (-0.02; *SE* = 0.06), while equality between the two estimates could not be rejected (*p* = .24). The cognitive tests and behavioral problems indices considered in Figure 5 were scored as the overall average of all corresponding index scores measured from age 5 to 14. We also considered age periods 5-6; 7-10; and 11-14 (see online appendix, Table S14-15). Head

Start impact on BPI index were stable across age groups and cohorts and were of similar magnitude as the 5-14 average.

[Insert Figure 5 here]

Deming (2009) reported some fadeout of Head Start impact on the cognitive tests index by ages 11-14: from an estimate of 0.15 *SD* (*SE* = 0.09) by age period 5-6 to one of 0.06 *SD* (*SE* = 0.06). In contrast, for the complement cohort, a fadeout from a small but positive estimate (0.06 *SD*; *SE* = 0.07) might have occurred faster by age period 7-10 (-0.03 *SD*; *SE* = 0.06): the difference in impact with Deming's cohort for this age group (0.13 *SD*, *SE* = 0.06) was of marginal significance ($p = .12$). Complement cohort estimates ended at -0.05 *SD* (*SE* = 0.07) by age period 11-14. Finally, the combined cohorts sample faced a similar trend as the complement cohort; overall, impacts approached zero earlier for later Head Start cohorts.

In our analytic model (see Equation 1), other 'Preschool' was also included as a within-family predictor of adult outcomes. Considering later and combined cohorts patterns of school age outcomes by age period: for both Head Start and other 'Preschool', impacts on cognitive outcomes were positive at treatment outset (age 5 to 6), followed by a fadeout, possibly occurring earlier for Head Start participants (see online appendix, model (5) in Table S14). Second, impacts on the nontest score index were unfavorable and statistically significant, for both preschool statuses (online appendix, Table S10). Third, impacts on the Behavior Problems Index were sometimes significant, mostly similar in magnitude, and unfavorable over all age periods across both preschool groups (online appendix model (5) in Table S15). Overall, we could never statistically reject the equality of estimates between Head Start and other 'Preschool' status for any of the considered school age and adulthood outcomes for the later cohorts of siblings (online appendix, Tables S10-S13 respective top panels).

**Robustness Checks**

As noted above, while the FFE design has been a workhorse empirical strategy to estimate the causal impact of Head Start (Currie and Thomas, 1995; GTC, 2002; Deming, 2009; Bauer and Schanzenbach, 2016), recent research has called into question this approach both in terms of the external (MSG, 2019) and construct validity of the FFE design (Heckman & Karapakula, 2019). The following section will discuss these threats and how, if at all, our primary findings change as a result.

**Selection into Identification.** MSG (2019) showed fixed effects (FE) can induce non-random selection of individuals into the FE identifying sample, leading to biased FE estimates relative to the ATE unless reweighting on observables is completed. In the FFE context, it may be that families with differential sibling participation in Head Start are systematically different across a variety of measures compared to those families with siblings that do not have variation in preschool status. If present, this "selection into identification" (SI) is a threat to the external validity of our results and could lead to a biased FFE estimate of Head Start's impact compared to the ATE (MSG, 2019).

To address this potential problem, following our above FFE analysis we performed the reweighting-on-observables procedure discussed in MSG (2019), first checking whether the combined cohort FFE identifying sample exhibited SI across a variety of observables including child's birth cohort, family size, mother's age at child's birth, permanent income, maternal AFQT, and whether a child was African American. Using the "switcher" and "non-switcher" naming convention found in MSG (2019)— where (non-)"switcher" represents those families with (no) sibling variation in preschool status—and estimating propensity scores via multinomial logistic regression, we found differences over all family characteristics between Head Start participating "switcher" families and non-Head Start "non-switcher" families. By contrast, the only statistically significant difference ($p < .001$) in predictors between Head Start "switcher" versus "non-switcher" families (i.e., in which all siblings attended the program) was the indicator whether a child was African American.

Given the FFE identifying sample for the combined cohort exhibited some degree of SI, we corrected the potentially biased FFE estimates using the one-step reweighting-on-observables procedure in MSG (2019). Overall, we found no evidence for the combined cohort that re-weighting changed the estimates of Head Start's impact on young adult outcomes including high school graduation and ASI. We also performed the MSG (2019) reweighting-on-observables procedure for the Deming cohort sample. Similar to MSG (2019), we found reweighting attenuated the FFE estimates of Head Start for high school graduation, idleness, learning disability and poor health, but find little change on other young adult outcomes or ASI. These results closely replicate MSG (2019) findings and can be found in Table S16 of the online appendix.

**Spillovers.** The construct validity of FFE estimates has also recently come into question. Revisiting the Perry Preschool Project, Heckman & Karapakula (2019) showed siblings who participated in the program had large positive spillovers on their non-participating siblings, particularly for male siblings. While this problem is material to the FFE design, it is of less concern for this paper since we were interested in comparing FFE results across cohorts and between short-run and long-run outcomes. In addition, it is unclear how sibling spillovers could explain this paper's main findings, unless sibling spillover effects had become much stronger in the later as opposed to earlier cohorts.

That said, given Head Start's provision of wraparound services, some of which may influence parenting practices, some degree of spillover across children is likely (Ludwig and Miller, 2007; Deming, 2009; Office of Head Start, 2019). GTC (2002) and Deming (2009) test for Head Start treatment spillovers by interacting an indicator for first born status with an indicator for Head Start treatment status. If spillover effects are present from an older to younger sibling, one would expect the impact of Head Start to be larger for non-first-born siblings (Deming, 2009). Consistent with Deming (2009), we found sparse and inconsistent evidence of spillovers. This holds true for both the Deming cohort and combined cohort samples (Table S17-S18).

**Reconciliation of Cross-Cohort Results**

**Human Capital Index.** One possible explanation for cohort differences in the estimated impacts of Head Start is that the more recent cohort was more advantaged, and thus less likely to benefit from Head Start. To check if this was the case, a household human capital factor was constructed (Cronbach's α = .83) by combining standardized measures of maternal, and both grandparents', education levels, maternal AFQT, the natural logarithm of family permanent income, and the CNLSY Home Observation Measurement of the Environment short-form (HOME). This human capital factor was then interacted with indicators 'Head Start' and 'Preschool' from equation (1). Interaction impact estimate on ASI, for the combined cohorts sample, was not statistically significant for any value within the range of the household human capital factor.

**Cohort Covariates**. To check whether the difference between Deming's cohorts and complement cohorts was due to some of the covariates we presented above, we interacted Head Start with an indicator for whether siblings belong to the complement cohort, along with interactions of Head Start with a series of covariates: first interacting one covariate at a time with the main effect included, then interacting all covariates with all main effects included. These covariates included: pre-treatment index; family human capital index; mother's age at child's birth; child's age at outcome measurement; indicators for gender, whether white/Hispanic or black; and whether maternal AFQT score was one *SD* below the mean. In addition, the 2007-2009 Great Recession could have negatively impacted complement cohort siblings who lived through its aftermath as teenagers. Thus, an indicator created for complement cohort siblings between age 12 and 18 in 2008 was also added. Had any of these interactions substantially reduced the estimate from Head Start-Cohort interaction, then these covariates would have explained some of the cross-cohort change in Head Start impacts between the Deming and complement cohort. We did not find any evidence that this was the case (Table S19).

**Blinder-Oaxaca Decomposition**. A Blinder-Oaxaca decomposition approach (Blinder, 1973; Oaxaca, 1973) allowed us to consider how much a group mean difference on an outcome of interest ($Y$) is explained by group differences in predictors. Using the Deming cohort group ($d$) and complement cohort group ($c$), we formalized a threefold decomposition (Jann, 2008) as:

$$Yd - Yc = E + C + I$$

where $E$ (= $Ed - Ec$) measures the part of complement cohort's expected change in ASI, with complement cohort's predictors' means (i.e., endowments) fixed at Deming's cohort levels; $C$ (= $Cd - Cc$) measures the part of complement cohort's expected change in ASI, with complement cohort coefficients fixed at Deming's cohort levels; and $I = (Ed - Ec)$ x $(Cd - Cc)$ measures the contribution of the interaction between endowments and coefficients respective cross-cohort differences. The results from this analysis highlight key differences between Deming's and complement cohort samples, and showed how these differences – most notably in the pre-treatment index and mother's age at child's birth – drive variation in estimated Head Start impacts across cohorts.

Breaking down the threefold decomposition results, we first found that the mean difference in ASI for Head Start attendees between the Deming cohort and complement cohort ($Yd - Yc$) was -0.22 SD (SE =.09).[26] The direction of this mean difference favored the complement cohort and was statistically significant. Second, having chosen the pre-treatment index and mother's age at child's birth as predictors in the analysis,[27] we found that predictors' endowment parts of the decomposition ($E$) explained all of the outcome group mean difference (-0.27 SD; SE = .10), with mother's age at child's birth endowment recovering the near totality of it (-0.21 SD; SE = .10). The coefficient ($C$) and interaction ($I$) parts of the decomposition were negligible and not statistically significant (p >.84; p >.95). In sum, with mother's age at child's birth fixed at Deming's cohort level, complement cohort Head Start attendees would share an ASI expected value similar to that of their Deming's cohort counterparts.

Further, if for siblings of the complement cohort's counterfactual non-preschool group the expected change in ASI mean was also explained by mother's age at child's birth (either as endowments, coefficients, or interaction effect), then, keeping this factor equal, Head Start would have had an impact of similar magnitude across cohorts. Thus, a threefold decomposition was conducted for counterfactual non-preschool group siblings. Cross-cohort mean difference on the outcome of interest ($Yd - Yc$) was moderate and statistically significant (-0.56 SD; SE = .09). However, the $E$ and $C$ components of the decomposition were negligible (p >.88; p > .74), whereas the $I$ component—i.e., the interaction of mother's age at child's birth endowments' and coefficients' differences—recovered the outcome differential (-0.61 SD; SE = .18).

Because of the NLSY design, later cohorts of children had, on average, older mothers. Earlier cohorts (i.e., Deming cohort) were born to relatively younger mothers, many of which may have disproportionately benefited from Head Start due to the program's provision of wraparound services that emphasize parental involvement (Currie and Neidell, 2007; Deming, 2009). This factor alone might have contributed to the discrepancy in our estimated Head Start impacts between Deming's and later cohorts. In particular, within Head Start participating families, siblings who did not attend a preschool program appeared to have on average benefitted the most from having an older mother during their early years. Finally, the explanatory power of the mother's age at child's birth predictor remained robust after adding other covariates to the Blinder-Oaxaca decomposition, including permanent family income, whether the mother attended college (Table 2), a family human capital index and family size (Table S19, notes).

**Secular Trend.** Our original cohort analysis looked only at the impact of Head Start across two cohorts – the Deming cohort and the complement cohort – obscuring whether over time there was a gradual secular decline or precipitous drop in the effect of Head Start. To address this, we decomposed our overall sample into three new birth-cohort-year groupings: C1 – families with all siblings born before 1983, C2 – 1983-1987 and C3: post–1987. More than 90% of the C1 siblings and around a third of the C2 siblings were part of the Deming cohort sample. Two thirds of the C2 siblings and all of C3 siblings were part of the complement cohort sample.

Across virtually all of our outcomes, Head Start more favorably impacted the C1 cohort compared with C2 and C3 cohorts. Moving from older to more recent cohorts, we observed unfavorable sign changes in the direction of Head Start's impact for a variety of short-term and longer-term outcomes including: cognitive and non-cognitive school age measures, ASI, educational attainment and earnings.[28] In addition, the estimated trends indicate a comparatively larger decline in adulthood outcomes for individuals belonging to the most recent post-1987 cohort. In contrast, for school-year nontest outcomes the decline was observed in much earlier cohorts. As reported in the online appendix Table S20, Head Start impacts between C1 and C3 on ASI went from positive and statistically non-significant (0.09 *SD*; *SE* = 0.09) to negative and significant at the 10% level (-0.25 *SD*; *SE* = 0.14). By contrast, Head Start impact on the nontests index went from a statistically significant 0.29 SD (*SE* = 0.10) advantage for C1 to a negligible -0.06 *SD* (*SE* = 0.15) for C3, with the drop occurring between C1 and C2 (-0.10 *SD*; *SE* = 0.15).

## Discussion and Conclusion

In this study we replicated and extended Deming's (2009) evaluation of Head Start impacts over life-cycle skill formation. We found mixed results for Deming's cohort of siblings, after having extended adulthood individual outcomes with an additional decade of CNLSY data. Second, replicating Deming's analytic framework on children born to CNLSY mothers after the children in Deming's cohort revealed contrasting patterns of impacts. In general, for this new cohort, impacts were negative. In fact, for our study of more recent cohorts, Head Start participation might have been detrimental, relative to home care, on non-cognitive and behavioral measures or on the adulthood summary index. Third, combining both cohorts produced Head Start impact estimates on all measured outcomes that were small and not statistically significant.

Deming (2009) estimated a 0.11 log points Head Start impact on adult wages. However, since survey participants were too young at the time of the calculation to report actual wages, this estimate was based on projected future adult wages. Given the benefit of time and 10 additional years of NLSY data, we estimate the impact of Head Start on observed adult wages for the Deming cohort. This was done for siblings 25 years-old or more, unconditional on employment and averaged across each survey-round up to 2014. We found a non-statistically significant smaller impact at 0.07 log points ($SE = 0.12$). Using dollar (2014 USD) instead of log earnings produces a negative and still non-statistically significant impact of -$999 ($SE = $1,507$). In sum, Head Start generated no clear adult earnings gain for the Deming (2009) cohort siblings, although the large confidence interval on our estimate includes Deming's original estimated impact.[29]

Johnson and Jackson (2019) analyzed earlier cohorts of siblings born before 1976 with a dynamic complementarity design (i.e., capitalizing on two exogenous sources of variation separated in time). They found a larger, more precise estimate: attending Head Start at age 4—i.e., facing an average Head Start spending versus no spending, *coupled* with (and sensitive to) an average public K-12 spending—boosted earnings of poor children (measured from age 20 to age 50) by 0.10 log points ($SE = 0.02$).[30] These positive estimates, generated from a different identification strategy and earlier cohorts, are close to Deming's projections, but the variation and imprecision associated with our estimates, underscores the importance of monitoring cross-cohort trends in Head Start impacts.

By and large, our analysis showed heterogeneous impacts from a large-scale early childhood education program. Changes in counterfactual conditions in household characteristics (e.g., maternal age) predicted cross-cohorts differences on long-term outcomes. Further, a substantial increase in spending on means-tested spending between the 1980s and 2010s, and possible changes to program quality (e.g., due to a steep and continuous enrollment increase) could also help explain these variations (Ludwig & Phillips, 2008), as could changes in labor market conditions and the return to specific forms of human capital over time. Changes in human capital returns over time may have induced individuals from more recent birth cohorts to invest less in educational attainment relative to earlier older birth cohorts. Our findings that Head Start negatively impacted 'Idleness' (i.e., not employed or enrolled in school) and 'Some College Attended' for more recent birth cohorts but not for earlier cohorts lends support to this claim.

Overall, this paper suggests that understanding and eliciting pathways of early skill formation with potential subsequent complementarities could be an important priority for basic human capital research and education policies. The novelty of these findings, combined with the possibility of unobserved changes in the selection process into Head Start during this period, necessitate further research on recent cohorts of Head Start attendees using complementary identification strategies.

**Notes**

1. An overview of the literature on Head Start's short and medium-term impacts is available in the online appendix S1.

2. For example, Johnson and Jackson (2019) calculated that for a child attending Head Start, a 10 percent increase in K-12 spending boosted educational attainment by 0.4 years, earnings by 20.6 percent; and reduced the probability of being incarcerated by 8 *pp*.

3. Nearly 95% of children included in the Deming (2009) family fixed effects estimation sample were born between 1976 and1986. The remaining 5% of children were born between 1970-1975.

4. The adulthood summary index used in Deming (2009) included high school graduation, college attendance, teen-age parenthood, idleness (i.e., neither working nor attending school), crime and poor health status.

5. The FFE analysis sample used in this paper includes 1970-1996 CNLSY birth cohorts.

6. Earnings are computed as the log of 1994-2014 averaged earnings, adjusted for age and survey-year.

7. The Head Start impact estimates of Carneiro and Ginja (2014), Bauer and Schanzenbach (2016), and Barr and Gibbs (2019) were based on samples that included 1990s birth cohorts. However, these studies did not systematically estimate Head Start impacts by birth cohorts across time. Instead primary results were for analyses based on overall samples which include both more recent birth cohorts from the late 1980s and early 1990s, and older birth cohorts from 1970s and early to mid-1980s.

8. A third restriction was applied to the complement cohort: siblings were considered for eligibility up to 2000 excluding those already part of Deming's cohort (i.e., selected under rule 1 and 2); it is in that sense that this new cohort is the complement of Deming's cohort. Of all siblings comprising the complement cohort, 78 percent had reached four years of age post 1990 (75 percent, for Head Start participants). As in Deming (2009) and for all cohorts, the original NLSY79 oversample of low-income whites was excluded. One final point, the sum of Deming's cohort and the complement cohort is smaller than the size of the combined cohorts sample as there are more opportunities for siblings to meet the family fixed effects eligibility criteria for the latter. For example, there were cases where families included in Deming's cohort had an additional child or children in later years that were not age-eligible for Deming's cohort but were old enough to be a candidate for the complement cohort. However, to be included in the complement cohort these age-eligible children needed to exhibit differential participation in Head Start. Thus, the family with one child was automatically disqualified from the complement cohort, and the family with multiple new children would also be excluded unless these children exhibited differential participation in Head Start. Regardless of their inclusion within the complement cohort, the children in each of these scenarios would be included in the combined cohort sample.

9. NLSY79 derived, from the Armed Services Vocational Aptitude Battery of tests, the Armed Forced Qualification Test (AFQT) comprising items in arithmetic reasoning; mathematics knowledge; word knowledge; and paragraph comprehension.

10. Deming (2009) presented these characteristics over three preschool statuses (i.e., Head Start; other preschool; no preschool) and by racial/ethnic subgroups. To keep Table 1 manageable, only overall means for Head Start and no preschool status (the counterfactual) are displayed. For details with other preschool status included (and also by racial/ethnic subgroups) see online appendix Tables S2-S4. There was a reduction in sample sizes when restricting on families who made difference choices of preschool status for at least 2 siblings (see data subsection, rule 2): by 66 percent for Deming's cohort, by 41 percent for the complement cohort; and by 45 percent for the combined cohorts. However, variation on selected household characteristics appeared to be very similar across both type of samples, and across all cohorts (Table 1).

11. Differences were even more pronounced when comparing between Head Start and the other "Preschool" status (see online appendix Table S2).

12. Both variables were derived from CNLSY cross-round item asking respondents which highest grade they had completed at the date of the latest survey round interview. Responses were recoded as equivalent years of completed schooling (e.g., if respondent answered "high school graduate" it was recoded 12; 'completed an associated degree' was recoded 14; etc.).

13. We opted to combine all respondent age 35 + into a single birth cohort dummy to ensure a sufficient sample size (N = 589) before regressing for adjustment. For all other birth cohorts, sample sizes were of at least 300 observations. The arguably arbitrary 35-year threshold was chosen such that *a priori* valid inputs would be available.

14. As in Deming (2009), when estimating Head Start impacts in the regression models, missing data for these covariates were imputed with corresponding sample mean value. For each, a dichotomous indicator for imputed responses was also included.

15. As in Deming (2009), variables comprising the pre-treatment covariates index were all first positively oriented with respect to the adulthood summary index. For example, variables like gender (male), age (older), or grand-mother living in household between child's birth and age 3, were negatively correlated with the outcome. Their correlational direction was reversed multiplying their sign by -1. All covariates were then standardized and aggregated into an index, in turn, also standardized (mean = 0; $SD$ = 1).

16. Deming (2009; pp. 123-124) used a similar approach in estimating impacts on the school age cognitive tests index.

17. From similar trends, Deming (2009) concluded—based on Altonji, Elder and Taber (2005) seminal work on the topic—that estimates obtained from model (3) stood as lower bounds for Head Start causal impacts.

18. Throughout (see online appendix, Tables S10-S13), we considered identical demographic subgroups as in Deming (2009). Looking at online appendix Table S10 'Adulthood index' column, for the combined cohorts sample most impacts of Head Start were also close to zero. One exception being for siblings with low maternal AFQT subgroup—one $SD$ below NLSY79 AFQT empirical sample average (online appendix Table S10, bottom panel). For these siblings (N = 810), Head Start appeared to have had a marginally significant positive impact on the adulthood summary index (0.11 $SD$; $SE$ = 0.08). This estimate was greater for Deming's cohort (0.38 $SD$, significant at the 1 percent level). For the complement cohort, impact was much smaller (0.03 $SD$, $SE$ = 0.15), although difference testing between this cohort's estimate and that of Deming's cohort did not fall below the 10 percent level of statistical significance ($p$ = .15). For the complement cohort, the proportion of siblings with low maternal AFQT background was also smaller (0.21 compared with 0.41 within Deming's cohort) which is consistent with the overall observed favorable shift over household characteristics between the two cohorts (see Table 1).

19. For all estimates on individual outcomes, overall and by subgroups, see online appendix tables S11-S12.

20. Education data were obtained using CNLSY 2014 survey-round *cross-round* variable for respondents' highest grade completed.

21. For both Deming's and complement cohorts, Head Start impacts on 'Poor health status' were positive (i.e., not self-identifying as being of poor health). This is in line with results found on a range of health outcomes in Carneiro and Ginja Head Start evaluation study (2014). The estimate for the combined cohorts was very small though, with a standard error well balanced across zero (Figure 3; online appendix Table S12).

22. Although we could not reject equality between these estimates, we detected a statistically significant Head Start favorable impact difference (see online appendix, Table S12) of about 8 *pp* between genders for the 'Idle' individual outcome (i.e., neither working nor in school) for the combined cohorts sample. Similarly, for the complement cohort, females had a 14 percentage-points advantage on the 'Crime' outcome (i.e., whether involved with the justice system).

23. Head Start impacts were positive and statistically significant for the subgroup of siblings whose maternal IQ background was 1 *SD* below the mean: adulthood index (0.38 *SD*; *SE* = 0.12); educational attainment (0.45 years of schooling completed; *SE* = 0.23); and earnings (0.44 log-points; *SE* = 0.20). See online appendix Table S13.

24. We conducted this section's analysis as in Deming (2009) and considered identical outcomes and age groups. The full set of estimates were compiled in the online appendix Tables S12-S13 and S19-S20.

25. Other 'Preschool' impact estimates had similar trends on all these 'noncognitive' outcomes (see online appendix Tables S12-S13).

26. In this complementary analysis, standard errors were clustered at the family level.

27. These predictors were selected based on both the magnitude and direction of the mean covariate differences between the Deming and complement cohort. Mother's age at child's birth was around 7 years higher, and the mean pre-treatment index statistically more favorable (0.20 SD; $SE = .08$) for complement cohort's Head Start attendees.

28. We also checked for impacts on an alternative earnings variable, taking this time the natural logarithm of the most recent yearly earnings available in CNLSY 2014 survey-round. Head Start impacts were never statistically significant.

29. Earnings (log transformed or not) regression estimates were very similar—0.05 log points; $SE = 0.11$; -1030 2014 USD; $SE = 1488$, respectively—whether age was controlled for instead of year of birth. As described in the results section, for Deming's cohort, attending Head Start versus no preschool yielded 0.3 years increase of completed schooling. From this, and based on Card's (1999) review on returns to education of about 5 to 10 percent per year of completed schooling, we might have expected to find evidence of an impact on earnings between 1 and 3.5 percent.

30. That estimate appeared to be sensitive to subsequent K-12 spending level: coupled with a 10% decrease in K-12 spending, the estimate fell to 0.03 log-point ($SE = 0.03$) and was no longer statistically significant at the 10 percent level. In contrast, with a 10% increase in K-12, the estimate jumped to 0.17 log-point ($p < .01$).

References

References

Anderson, K., Foster, J., & David Frisvold, D. (2010). Investing in Health: The Long-Term Impact of Head Start on Smoking. *Economic Inquiry 48*(3), 587-602.

Anderson, M. L. (2008). Multiple inference and gender differences in the effects of early intervention: A reevaluation of the Abecedarian, Perry Preschool, and Early Training Projects. *Journal of the American statistical Association*, *103*(484), 1481-1495.

Bailey, M. J., Sun, S., & Timpe, B. (2018). Prep School for Poor Kids: The Long-Run Impacts of Head Start on Human Capital and Economic Self-Sufficiency. Unpublished manuscript, Department of Economics, University of Michigan, Ann Arbor, MI.

Barr, A., & Gibbs, C. R. (2019). *Breaking the Cycle? Intergenerational Effects of an Anti-Poverty Program in Early Childhood*. Manuscript submitted for publication.

Bauer, L., & Schanzenbach, D.W. (2016). The Long-term Impact of the Head Start Program. Policy report. Retrieved from The Hamilton Project website: http://www.hamiltonproject.org/papers/the_long_term_impacts_of_head_start?_ga=2.58041645.688673869.1554397866-1708819741.1554397866

Blinder, A. (1973). Wage Discrimination: Reduced Form and Structural Estimates. The Journal of Human Resources, *8*(4), 436-455.

Card, D. (1999). The causal effect of education on earnings. In *Handbook of labor economics* (Vol. 3, pp. 1801-1863). Elsevier.

Carneiro, P., & Ginja, R. (2014). Long-term impacts of compensatory preschool on health and behavior: Evidence from Head Start. *American Economic Journal: Economic Policy*, *6*(4), 135-73.

Chetty, R., Friedman, J. N., Hilger, N., Saez, E., Schanzenbach, D. W., & Yagan, D. (2011). How does your kindergarten classroom affect your earnings? Evidence from Project

STAR. *The Quarterly Journal of Economics*, *126*(4), 1593-1660.

Cunha, F., Heckman, J. J., Lochner, L., & Masterov, D. V. (2006). Interpreting the evidence on life cycle skill formation. *Handbook of the Economics of Education*, *1*, 697-812.

Currie, J., & Almond, D. (2011). Human capital development before age five. In *Handbook of labor economics* (Vol. 4, pp. 1315-1486). Elsevier.

Currie, J., & Neidell, M. (2007). Getting Inside the "Black Box" of Head Start Quality: What Matters and What Doesn't. *Economics of Education Review*, *26*(1), 83-99.

Currie, J., & Thomas, D. (1995). Does Head Start Make a Difference? *The American Economic Review*, *85*(3), 341-364.

Currie, J., & Thomas, D. (1999). Does head start help Hispanic children?. *Journal of public Economics*, *74*(2), 235-262.

Deming, D. (2009). Early childhood intervention and life-cycle skill development: Evidence from Head Start. *American Economic Journal: Applied Economics*, *1*(3), 111-34.

Duncan, G. J., & Magnuson, K. (2013). Investing in preschool programs. *Journal of Economic Perspectives*, *27*(2), 109-32.

Frisvold, D. E. (2006). Head Start participation and childhood obesity. *SSRN Electronic Journal*.

Frisvold, D. E., & Lumeng, J. C. (2011). Expanding exposure: Can increasing the daily duration of head start reduce childhood obesity? *Journal of Human resources*, *46*(2), 373-402.

Garces, E., Thomas, D., & Currie, J. (2002). Longer-term effects of Head Start. *American economic review*, *92*(4), 999-1012.

Heckman, J. J., & Karapakula, G. (2019). Intergenerational and Intragenerational Externalities of the Perry Preschool Project. *NBER Working Paper Series*, 25889.

Heckman, J. J., & Mosso, S. (2014). The economics of human development and social mobility.

Annual Review of Economics, 6(1), 689-733.

Heckman, J. J., & Pinto, R. (2015). Econometric mediation analyses: Identifying the sources of treatment effects from experimentally estimated production technologies with unmeasured and mismeasured inputs. *Econometric reviews*, 34(1-2), 6-31.

Hoynes, H., Schanzenbach, D. W., & Almond, D. (2016). Long-run impacts of childhood access to the safety net. *American Economic Review*, 106(4), 903-34.

Jann, B. (2008). The Blinder–Oaxaca Decomposition for Linear Regression Models. *The Stata Journal: Promoting Communications on Statistics and Stata*, 8(4), 453-479.

Johnson, R., & Jackson, C. (2019). Reducing Inequality through Dynamic Complementarity: Evidence from Head Start and Public School Spending. *American Economic Journal: Economic Policy*, 11(4), 310-349.

Ludwig, J., & Miller, D. L. (2007). Does Head Start improve children's life chances? Evidence from a regression discontinuity design. *The Quarterly journal of economics*, 122(1), 159-208.

Ludwig, J., & Phillips, D. A. (2008). Long-term effects of Head Start on low-income children. *Annals of the New York Academy of Sciences*, 1136(1), 257-268.

Miller, D. L., Shenhav, N. A., & Grosz, M. Z. (2019). Selection into Identification in Fixed Effects Models, with Application to Head Start. *NBER Working Paper Series*, 26174.

Oaxaca, Ronald. (1973). Male-Female Wage Differentials in Urban Labor Markets. *International Economic Review*, 14(3), 693-709.

Thompson, O. (2018). Head Start's Long-Run Impact Evidence from the Program's Introduction. *Journal of Human Resources*, 53(4), 1100-1139.

U.S. Department of Health and Human Services, Administration for Children and Families.

(2018). *Head Start Program Information Reports.* Retrieved from:

https://hses.ohs.acf.hhs.gov/pir/

Table 1. Household Characteristics Averaged over Head Start and No Preschool Status

| | Head Start | | | No preschool | | | Diff. HS-None | Diff HS-All |
|---|---|---|---|---|---|---|---|---|
| | Deming's cohort | Complement cohort | Combined cohorts | Deming's cohort | Complement cohort | Combined cohorts | | |
| *Permanent income* | | | | | | | | |
| Full sample | 32,884 [21,810] | 39,800 [27,539] | 35,970 [24,830] | 42,764 [30,000] | 61,857 [49,704] | 52,445 [41,989] | -0.37/-0.49/-0.44 | -0.50/-0.47/-0.46 |
| Fixed effects subsample | 34,672 [25,443] | 40,465 [29,118] | 37,571 [26,961] | 41,587 [27,968] | 61,793 [53,019] | 53,938 [45,831] | -0.26/-0.45/-0.41 | -0.35/-0.53/-0.48 |
| *Mother < high school* | | | | | | | | |
| Full sample | 0.24 [0.43] | 0.12 [0.33] | 0.19 [0.39] | 0.24 [0.43] | 0.12 [0.32] | 0.18 [0.38] | 0.01/0.01/0.02 | 0.15/0.14/0.18 |
| Fixed effects subsample | 0.28 [0.45] | 0.14 [0.35] | 0.20 [0.40] | 0.22 [0.42] | 0.12 [0.33] | 0.16 [0.37] | 0.12/0.06/0.11 | 0.23/0.15/0.21 |
| *Mother some college* | | | | | | | | |
| Full sample | 0.28 [0.45] | 0.40 [0.49] | 0.33 [0.47] | 0.25 [0.43] | 0.44 [0.50] | 0.34 [0.48] | 0.08/-0.09/-0.02 | -0.07/-0.28/0.21 |
| Fixed effects subsample | 0.25 [0.43] | 0.42 [0.49] | 0.34 [0.47] | 0.27 [0.44] | 0.43 [0.50] | 0.37 [0.48] | -0.04/-0.03/-0.06 | -0.13/-0.14/-0.15 |
| *Maternal AFQT* | | | | | | | | |
| Full sample | -0.61 [0.61] | -0.50 [0.71] | -0.56 [0.66] | -0.36 [0.80] | 0.11 [1.03] | -0.12 [0.95] | -0.33/-0.63/-0.50 | -0.47/-0.78/-0.66 |
| Fixed effects subsample | -0.62 [0.61] | -0.48 [0.70] | -0.54 [0.66] | -0.21 [0.81] | -0.01 [0.99] | -0.18 [0.92] | -0.31/-0.52/-0.43 | -0.40/-0.58/-0.50 |
| *Grandmother's education* | | | | | | | | |
| Full sample | 9.16 [3.09] | 9.69 [3.08] | 9.39 [3.09] | 9.45 [3.23] | 10.41 [3.39] | 9.92 [3.35] | -0.09/-0.22/-0.16 | -0.23/-0.37/-0.33 |
| Fixed effects subsample | 9.13 [3.12] | 9.79 [3.00] | 9.50 [3.04] | 9.69 [3.14] | 10.29 [3.30] | 10.07 [3.22] | -0.18/-0.16/-0.18 | -0.23/-0.22/-0.24 |
| Sample size | 779 | 637 | 1,491 | 1,931 | 1,857 | 3,658 | | |
| Sample size FE | 435 | 475 | 972 | 769 | 1,098 | 1,799 | | |

*Notes.* Means and standard deviations were obtained for the full and sibling fixed effects samples, across cohorts. Differences in means (in *SD* units) between *Head Start* vs. *No preschool* status (Difference HS-None) and between *Head Start* vs. *No preschool + Other preschool* status (Difference HS-All) were reported for Deming's cohort/Complement cohort/Combined cohort (in that order). Permanent income is the average over reported years of household net income (in 2014 dollars). The AFQT was age normed based on the NLSY79 empirical age distribution of scores, then standardized (mean = 0; *SD* = 1). Household characteristics for the *Other preschool* status group were documented in the online appendix, Table S2. Mean differences between Deming's cohort and Complement cohort, were all significant at the 1 percent level (with the exception of Grandmother's education; see online appendix Table S2).

Table 2. Head Start Impacts on Cohorts' Adulthood Summary Index

| | | Head Start | Other preschool | $p$-value (HS = Other) | $R^2$ |
|---|---|---|---|---|---|
| Deming (2009)[a] | (1) | **0.14** (0.07) | 0.08 (0.08) | .47 | 0.12 |
| Measurement period[b]: 1994-2004 | (2) | **0.27** (0.08) | 0.11 (0.08) | .12 | 0.59 |
| Sample size = 1,251 [364/364] | (3) | **0.23** (0.07) | 0.07 (0.07) | .08 | 0.62 |
| Deming's cohort[c] | (1) | **0.14** (0.06) | 0.08 (0.06) | .40 | 0.21 |
| Measurement period: 1994-2014 | (2) | **0.18** (0.07) | 0.02 (0.09) | .07 | 0.64 |
| Sample size = 1,251 [364/364] | (3) | **0.17** (0.07) | 0.03 (0.07) | .13 | 0.69 |
| Complement cohort[d] | (1) | **-0.12** (0.06) | 0.03 (0.06) | .01 | 0.24 |
| Measurement period: 2004-2014 | (2) | -0.16 (0.10) | -0.05 (0.08) | .30 | 0.71 |
| Sample size = 2,144 [497/795] | (3) | **-0.15** (0.07) | -0.04 (0.05) | .15 | 0.73 |
| $p$-value for model (3)[e] (Deming's = complement) | | .01 | .64 | | |
| Combined cohorts[f] | (1) | -0.02 (0.04) | 0.06 (0.03) | .05 | 0.24 |
| Measurement period: 1994-2014 | (2) | -0.01 (0.01) | 0.01 (0.04) | .78 | 0.61 |
| Sample size = 3,738 [951/1,275] | (3) | -0.01 (0.04) | -0.003 (0.04) | .86 | 0.63 |

*Notes:* Adulthood summary index (standardized) is a composite of 6 indicators: high school graduation; college attendance; teen-age parenthood; either working or attending school; involvement with the justice system; and poor health status. Model (1): adulthood index is regressed on Head Start and other preschool participation indicators, along with pre-treatment covariates and standardized permanent income; maternal AFQT score; one indicator for maternal high school graduation and one for some college attendance; siblings' gender and age. Model (2): same as model (1) but with family fixed-effect only, no pre-treatment covariates. Model (3): same as model (2) with pre-treatment covariates included. Standard errors are in parenthesis and clustered at the family level. Estimates in bold case were significant at the 5 percent level or less. [a] Deming published results. [b] Outcomes measurement period. [c] For Deming's cohort (compared with Deming 2009), individual outcomes composing the adulthood index were extended up to 2014. [d] Complement cohort includes siblings fitting the same criteria as in Deming (2009) but found eligible from 1990 to 2000. [e] $p$-value = estimates' difference testing between Deming's and complement cohorts' impacts estimated in model (3). [f] Combined cohorts integrates both Deming's and the complement cohorts.

Figure 1. Birth and Outcomes Time Range by Cohorts

*Notes:* Time-wise, the combined cohorts sample (not shown) encompasses Deming's and complement cohorts. Boundary end points are approximate, they include nonetheless the bulk of each distribution: around 95 and 85 percent for Deming's cohort and complement cohort, respectively.

Figure 2. Pre-Treatment Index Kernel Density Estimation by Preschool Status across cohorts

*Notes:* Distributions were smoothed, and densities estimated via the Epanechnikov kernel function. Kolmogorov-Smirnov tests all indicated non-equality, at the 0.1 percent level, between compared densities.

Figure 3. Head Start Impacts on the Adulthood Index Individual Outcomes Across Cohorts

*Notes*: Measurement period are displayed to right of each label. Impacts are expressed as proportions. Deming's cohort, N = 1,251; complement cohort, N = 2,144; combined cohorts, N = 3,768. The counterfactual was a no preschool attendance. Error bars represent standard errors which were clustered at the family level. Estimates were oriented such that a positive value represents a more favorable outcome.

Figure 4. Head Start Longer-Run Impacts

*Notes:* Measurement period are displayed to the right of each label. Impacts are expressed in standard deviation units. For both cohorts, N = 1,251. Recall that Deming (2009) study did not estimated Head Start impacts on educational attainment, college graduation and earnings; hence no Deming (2009) bar-estimates for these outcomes. The counterfactual was a no preschool attendance. Error bars represent standard errors which were clustered at the family level.

Figure 5. Head Start Impacts on School Age Outcomes Across Cohorts

*Notes:* Measurement period are displayed to the right of each label. Impacts are expressed in standard deviation units. Deming's cohort, N = 1,251; complement cohort, N = 2,144; combined cohorts, N = 3,768. The counterfactual was a no preschool attendance. Error bars represent standard errors which were clustered at the family level. Estimates were oriented such that a positive value represents a more favorable outcome. Deming (2009) estimated impacts on Behavioral problems index (not statistically significant) but did not report them. 'Grade retention' and 'Learning disability' composed the 'Nontests index'.

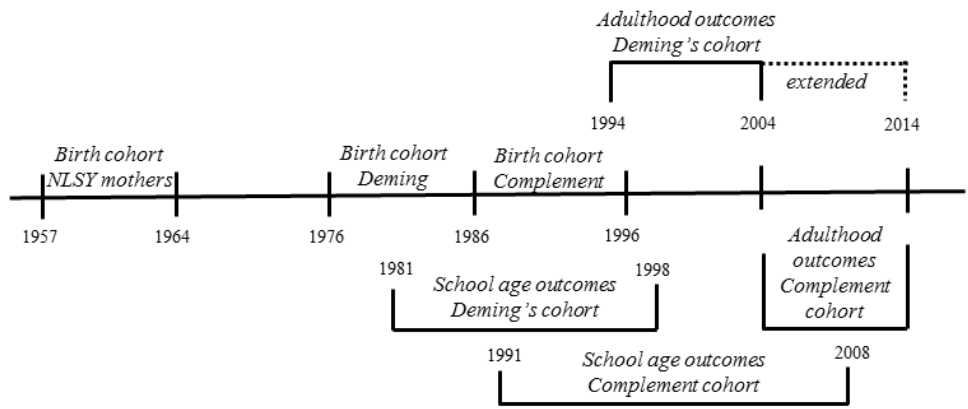

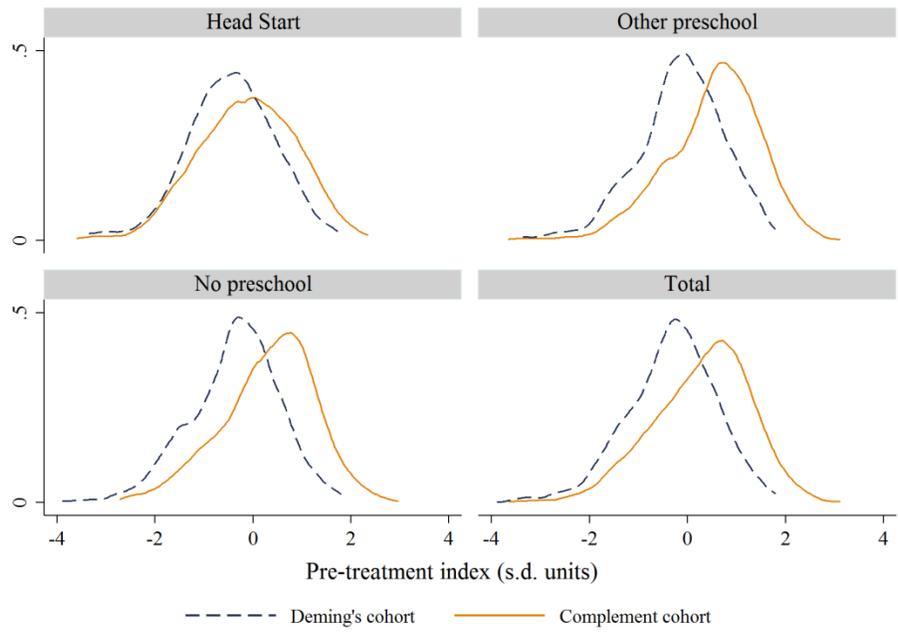

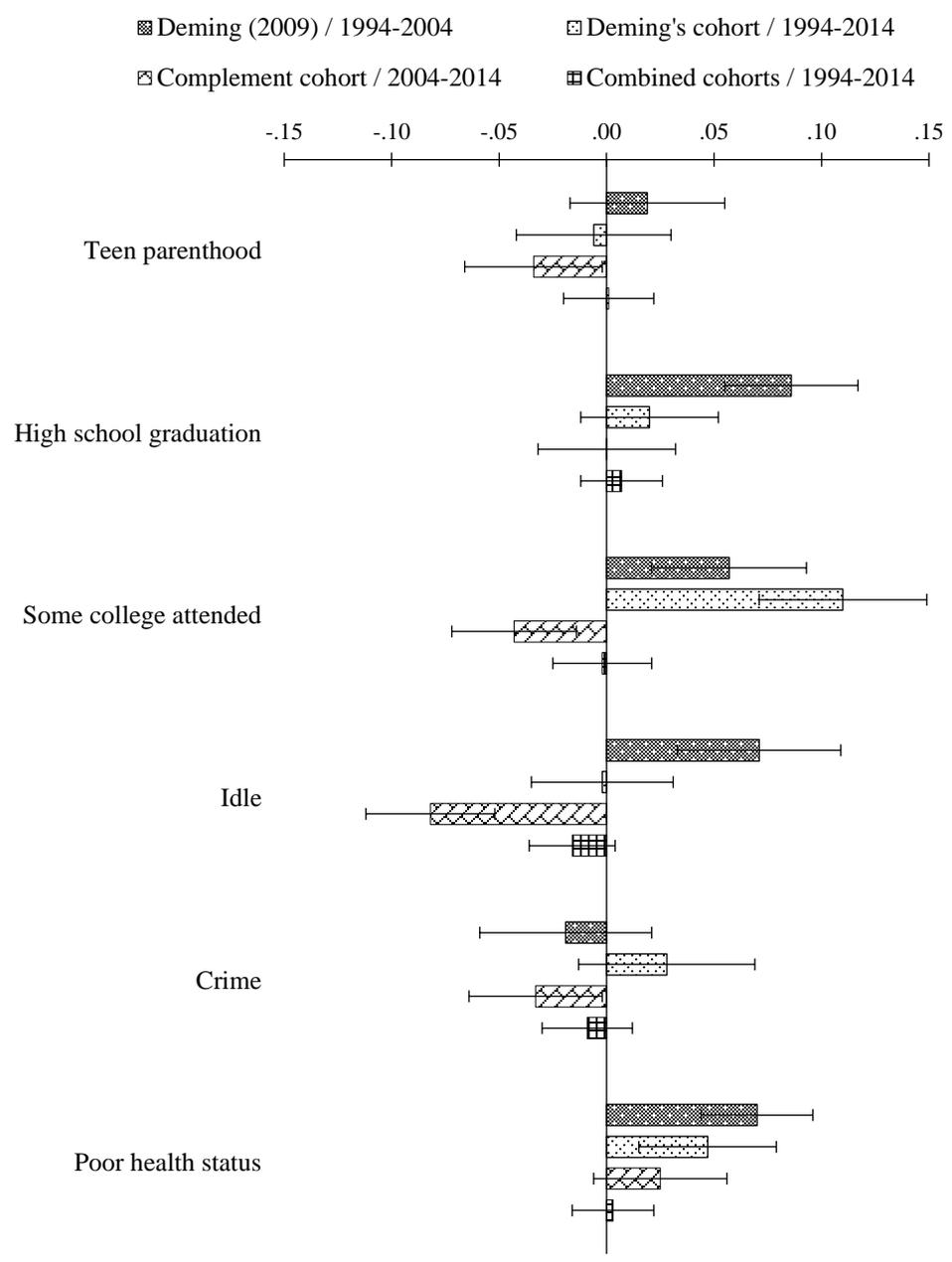

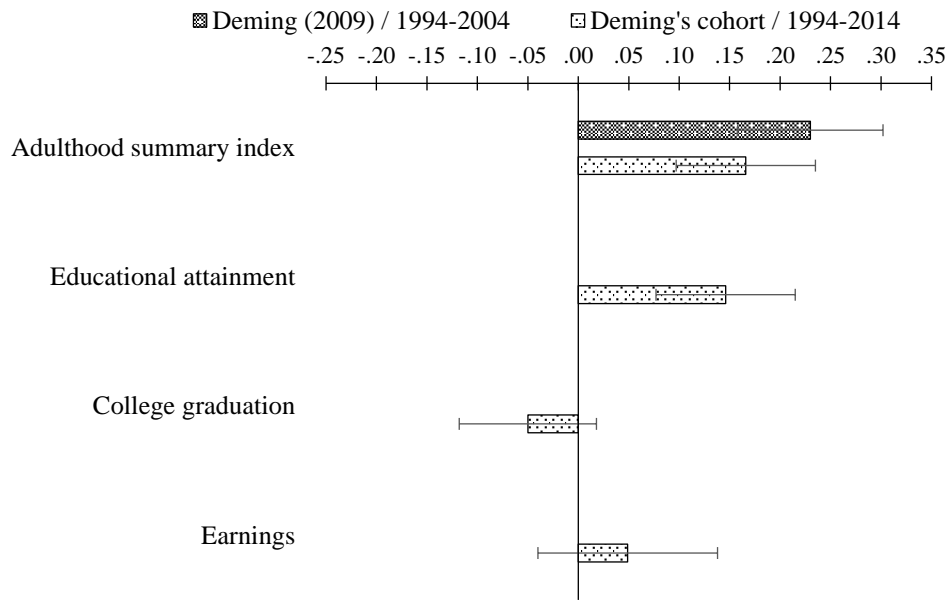

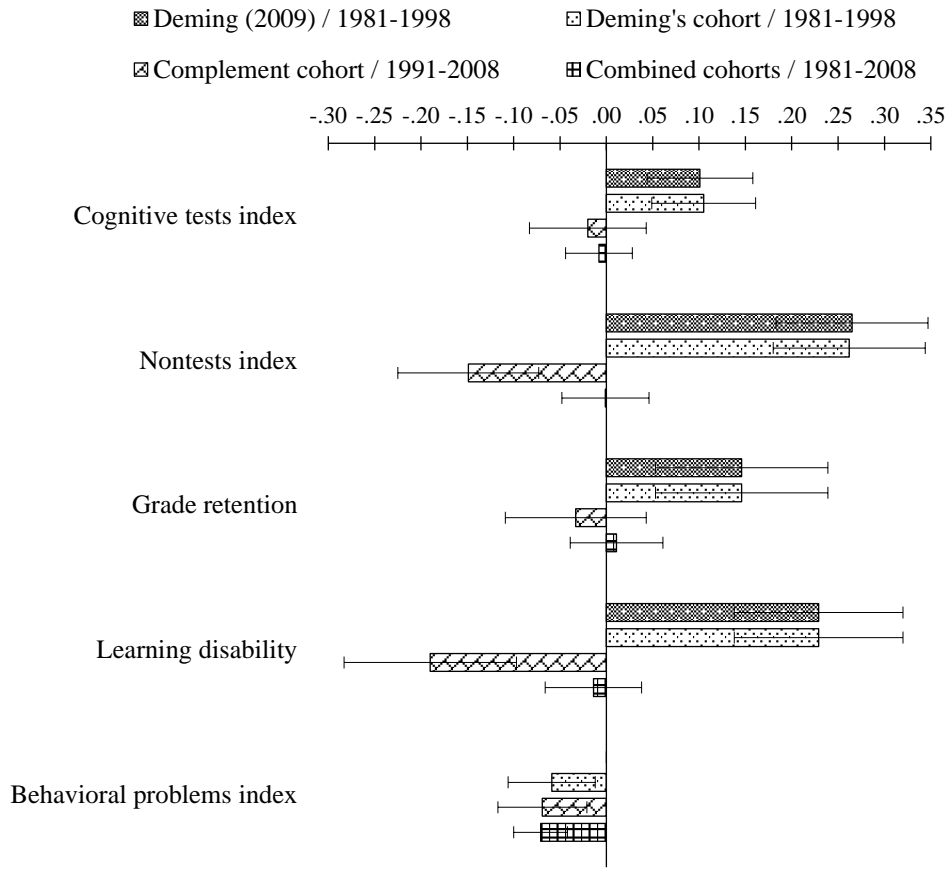

# Appendix

The following acronyms are used in this appendix:

**ACS** = American Community Survey

**AMTE** = Average Marginal Treatment Effect

**ATET** = Average Treatment-Effect-On-The-Treated

**BPI** = Behavioral Problems Index

**CAP** = Community Action Program

**CDS** = Child Development Supplement

**CESD** = Center for Epidemiological Studies Depression

**CNLSY** = Children of the NLSY 1979 cohort

**DID** = Difference-in-Difference

**FE** = Fixed Effects

**FFE** = Family Fixed Effects

**FOS** = Federal Outlays System

**GTC** = Garces, Thomas and Currie (2002)

**HS** = Head Start

**ICPSR** = Inter-university Consortium for Political and Social Research

**ITT** = Intent-to-Treat

**IV** = Instrumental Variables

**NARA** = National Archives and Records Administration

**NELS** = National Education Longitudinal Study

**NLSY** = National Longitudinal Survey of Youth, 1979 cohort

**OEO** = Office of Economic Opportunity

**PIAT** = Peabody Individual Achievement

**PIAT-M** = Peabody Individual Achievement Math

**PIAT-RR** = Peabody Individual Achievement Reading Recognition

**PIAT-RC** = Peabody Individual Achievement Reading Comprehension

**PPVT** = Picture Peabody Vocabulary Test

**RD** = Regression Discontinuity

**SAIPE** = Small Area Income and Poverty Estimates

**SEO** = Survey of Economic Opportunity

**SFR** = School Finance Reforms

**SEER** = Surveillance, Epidemiology, and End Results

**SSA** = Social Security Administration

TABLE A1. Selected Quasi-Experimental Studies of Head Start - Adolescence and Later Life Outcomes

| Study and Data | Study Design | Results |
|---|---|---|
| **FAMILY FIXED EFFECTS** | | |
| Does Head Start Make a Difference? (Currie and Thomas, 1995) Birth Cohorts: 1970-1986 Data includes NLSY mothers (born 1957-1965) and CNLSY children born to them from 1986, 1988 and 1990 sampling waves. FFE Sample (PPVT): N = 3,477. FFE Sample (Repeated Grade): N = 728. FFE Sample (Measles): N = 4,165. FFE Sample (Height by Age/Gender): N = 4,092. | FFE regression of outcome on controls for child age, gender, first born status, household income at the time child was age 3, mother FE, and indicators for HS participation and other pre-k participation. Focus on childhood outcomes which include PPVT score, grade repetition, measles immunization and height standardized by age and gender. | **African American:** HS (compared to no pre-k), no significant impacts on grade repetition, PPVT scores or height standardized by age and gender, +9.4% ($SE = 0.03$) more likely to get measles immunization. **White:** HS (compared to no pre-k) caused +47.3% ($SE = .12$) less likely repeat a grade, +5.9% ($SE = 1.52$) PPVT scores, +8.2% ($SE = 0.03$) more likely to get measles immunization. No significant impact on height standardized by age and gender. **Heterogeneity:** (1) African Americans: +6.8% ($SE = 1.93$) PPVT at base age, -1.3% ($SE = 0.31$) PPVT X Age interaction, no significant Repeat Grade X AFQT interaction. (2) Whites: +6.9 percent ($SE = 2.40$) PPVT at base age, no significant PPVT X Age interaction, +8.3% ($SE = 0.32$) less likely to repeat a grade for a +10% in normalized maternal AFQT. |
| Does Head Start Help Hispanic Children? (Currie and Thomas, 1999) Birth Cohorts: 1970-1988 Data includes NLSY mothers (born 1957-1965) and CNLSY children born to them from 1986, 1988, 1990 and 1992 sampling waves. Sample restricted to Hispanic children aged 5 and older at time of survey. Sample: N = 750. | FFE regression of outcome on controls for child's age, gender, first born status, presence of spouse/partner at age 3, mother employed at age 3 and ln(mean household income) (while child age 3-5), family FE, and indicators for HS participation and other pre-k. Outcomes include PPVT, PIAT-M and PIAT-RR test scores in addition to grade repetition. | **Overall:** HS (compared to no pre-k) participation caused +0.22 (t = 2.09) probability of not repeating a grade, +9.86 (t = 4.06) PPVT percentile score, +5.15 (t = 1.86) PIAT-M percentile score, +3.05 (t = 0.99) PIAT-RR percentile score. No significant impact of other pre-k (compared to no pre-k). Between Hispanic and non-Hispanic white children, HS closes 2/3 of the gap in probability of grade repetition and at least 1/3 of the gap in test scores. |

TABLE A1. (CONTINUED)

| Study | Method | Results |
|---|---|---|
| Longer-Term Effects of Head Start (Garces, Thomas and Currie, 2002) Birth Cohorts: 1966-1977 Individuals born 1966-1977 birth cohorts in PSID. Focus on adults age 18-30 by 1995. Oldest PSID birth cohorts excluded from sample. FE Sample (Panels A, B and D): N = 1,742. FE Sample (Panel C): N = 728. | FFE regression of outcome on controls for year of birth, race and gender, mother FE, and indicators for HS participation and other pre-k participation. Focus on adulthood outcomes which include completion of high school, attendance of some college, ln(earnings) if household member worked, and crime (ever reported booked or charged of a crime). | **Overall:** HS (compared to no pre-k) had no significant impact on high school completion, +9.2% ($SE = 0.05$) on attended some college, and no significant impact on ln(earnings) at age 23-25 and booked/charged with a crime. **Subgroup:** (1) African-Americans: no significant impact of HS on high school completion, attended some college, ln(earnings) age 23-25, -0.12 ($SE = 0.05$) less likely to be booked or charged with a crime. (2) Whites: +20.3% ($SE = 0.10$) more likely to complete high school, +28.1% ($SE = 0.11$) more likely to attend some college, no significant impact of HS on ln(earnings) age 23-25 and booked or charged with a crime. |
| Early Childhood Intervention and Life-Cycle Skill Development: Evidence from Head Start (Deming, 2009) Birth Cohorts: 1970-1986 Data includes NLSY mothers born 1957-1965 and CNLSY children born 1970-1986. Most children enrolled in HS from 1984-1990 and over 4 years old by 1990. FE Sample: N = 1,251. | FFE regression of outcome on controls gender, age and first-born status, pre-treatment covariates, mother FE, and indicators for HS participation and other pre-k participation. Outcomes include 1) short-term: PPVT, PIAT Math and Reading Recognition tests, grade retention and learning disability and 2) long-term: high school graduation, college attendance, idleness, crime, teen parenthood and self-reported health status, and a summary index of these young adulthood outcomes. | **Individual Outcomes:** HS participation (compared to no pre-k) impacted high school graduation +0.09 ($SE = 0.03$), no significant impact of college attendance, -0.07 ($SE = 0.04$) grade repetition, -0.06 ($SE = 0.02$) learning disability, -0.07 ($SE = 0.04$) idleness, no impact on crime or teen parenthood, -0.07 ($SE = 0.03$) poor health. **Test Score Index:** HS participation (compared to no pre-k) impacted test scores by +0.15 $SD$ ($SE = 0.08$) (ages 5-6), +0.13 $SD$ ($SE = 0.06$) (ages 7-10) and +0.06 $SD$ ($SE = 0.06$) (ages 11-14). **Nontest Index:** HS participation (compared to no pre-k) impacted non-cognitive +0.27 $SD$ ($SE = 0.08$), +0.35 $SD$ ($SE = 0.12$) (black), no significant impact for white/Hispanic), +0.39 $SD$ ($SE = 0.12$) (male), no significant impact for female. |

*(continued)*

TABLE A1. (CONTINUED)

| | | |
|---|---|---|
| | | **Adulthood Index:** HS participation (compared to no pre-k) impacted adulthood summary index by +0.23 *SD* (*SE* = 0.07), +0.24 (*SE* = 0.10) (black), +0.22 *SD* (*SE* = .10) (white/Hispanic), +0.18 *SD* (*SE* = 0.10) (male), +0.27 *SD* (*SE* = 0.11) (female) <br> **Fade-Out:** Large test score fadeout for blacks, none for white/Hispanic. |
| Investing in Health: The Long-Term Impact of Head Start on Smoking (Anderson, Foster and Frisvold, 2010) <br> Birth Cohorts: 1963-1978 <br> Sample (1999): N = 922. <br> Sample (2003): N = 1,005. <br> Data from PSID. Smoking data from 1999 and 2003 surveys for participants aged 21-36. | FFE regression of outcome on controls age, gender, race, birth order, first born status and a battery of family-level pre-treatment covariates, mother FE, and indicators for HS participation and other pre-k participation. <br> Primary outcome of interest is participation in smoking as a young adult (aged 21-36). | **1999 Sample**: HS (compared to no pre-k) participants -17.3% (*SE* = 0.08) less likely to smoke as a young adult. No significant impact of other pre-k (compared to no pre-k). -24.8% (*SE* = 0.10) difference in probabilities of smoking between HS and other pre-k. <br> **2003 Sample:** No significant impact smoking as a young adult for HS (compared to no pre-k) and other pre-k (compared to no pre-k). -19.4% (*SE* = 0.10) difference in probabilities in smoking between HS and other pre-k. Difference no longer statistically significant after controlling for educational attainment. <br> **Cost Benefit:** $9,967 PV of smoking reduction (3% discount rate). Avg. costs of HS participant in 2003 was $7,092. Value of smoking reduction 36-141% of HS program costs. |

*(continued)*

TABLE A1. (CONTINUED)

| | | |
|---|---|---|
| The Long-Term Impact of the Head Start Program (Bauer and Schanzenbach, 2016) Birth Cohorts: 1970-1990 Data includes NLSY mothers (born 1957-1965) and CNLSY children born to them. Analyzes HS treatment cohorts from 1974-1994. Two samples are used: 1) FE Long-Term: N = 1,439 individuals in 666 families and 2) FE Second Generation: N = 617 individuals in 300 families. Analysis sample limited to respondents 28 years or older in their most recent sample year and did not attrite after 2010 survey year. | FFE regression of outcome on controls for birth year fixed effects, gender, and pre-treatment characteristics. Age at response is also controlled for in self-control and self-esteem regressions. Although not explicitly provided in model details, it is assumed authors also included family fixed effect, indicator for HS and indicator for pre-k in their model specification given their indicated FFE strategy for uncovering HS impacts. Outcomes include high school graduation rates, some college, post-secondary education (i.e., license or certificate, associate's degree, or bachelor's degree), self-control index, self-esteem, and second-generation parenting practices index. | **Overall:** HS (compared to no pre-k) caused +5% ($p<0.1$) high school graduation rates, +12% ($p<0.1$) some higher education, +10% ($p<0.1$) post-secondary completion, +0.15 *SD* ($p<0.1$) self-control index, +0.15 *SD* ($p<0.1$) self-esteem index, +0.25 *SD* ($p<0.1$) positive parenting index. **Subgroup:** For children of mothers with less than high school education: +10% ($p<0.1$) high school graduation, +12% ($p<0.1$) some higher education, +0.3 *SD* ($p<0.1$) in self-control index, +0.25 *SD* ($p<0.1$) in self-esteem index. For African American children: +12% ($p<0.1$) some higher education, +8% post-secondary completion, +0.35 *SD* ($p<0.1$) self-control index, +0.20 *SD* ($p<0.1$) self-esteem index, +0.3 *SD* ($p<0.1$) positive parenting index. For Hispanic children: +10% ($p<0.1$) high school graduation, +12% ($p<0.1$) some higher education, +14% post-secondary completion. |
| Selection into Identification in Fixed Effects Models, with Application to Head Start. (Miller, Shenhav and Grosz, 2019) [NBER Working Paper] PSID Birth Cohorts: 1966-1987 CNLSY Birth Cohorts: 1970-1986 **PSID**: Individuals born 1966-1977 from PSID (replication of GTC), individuals born 1966-1987 from PSID with survey participants from SEO (extension of GTC). | FFE regression of adult outcome for a child on controls for individual and family level characteristics, mother FE, and dummy indicators for HS participation and pre-k participation. Weights are used to make sample representative of national population. Standard errors are clustered at mother level. | **PSID Replication:** no impact on high school graduation of HS participation, 0.15 (*SE* = 0.05) effect on some college (white males), no impact on crime, no impact on age 23-25 log earning. On whole, authors find replication matches GTC results. **PSID Extended:** HS leads to no significant improvements in high school completion, 0.12 (*SE* = 0.05) increase in the likelihood of attending some college for white children, no significant effect on reductions in criminal activity. No significant impacts on several summary indices of long-run |

TABLE A1. (CONTINUED)

| | | |
|---|---|---|
| Full Sample (replication): N = 3,399. FFE Sample (replication): N = 1,742. Full Sample (extended): N = 7,363. FFE Sample (extended): N = 5,361. FFE Sample (white / extended): N = 2,986. **CNLSY:** All children from CNLSY at least 4 years old by 1990. Identical sample as using in Deming (2009) and linked to NLSY mothers. | Adult outcomes include index of economic sufficiency, index of health outcomes, high school completion, criminal activity, college attendance. | economic and health outcomes or college completion. FFE is 50% larger than the AMTE for the representative sample (i.e., FFE produce LATE that overweights larger families) **CNLSY Replication:** HS (compared to no pre-k) lead to +8.5 *pp* in high school graduation (p<0.01), -7.2 *pp* in idleness (p<0.1), -5.9 *pp* in learning disability (p< 0.01) and -6.9 *pp* (p<0.01) in reporting poor health. **Reweighted CNLSY:** The reweighted estimate of HS (compared to no pre-k) lead to 40% smaller impact on high school graduation (p<0.1), 34% smaller impact on idleness, 4% smaller impact on disability and 25% smaller impact on poor health. Only reweighted estimates of high school graduation are statistically different than FFE estimates (p<0.1). |
| Breaking the Cycle? Intergenerational Effects of an Anti-Poverty Program in Early Childhood (Barr and Gibbs, 2019) [APPENDIX] [Working Paper] Child Birth Cohorts: 1970-1992. NLSY respondents from 1957-1965 birth cohorts and CNLSY children of these mothers. Sample restricted to individuals over 20 by 2012. FE Sample: N = 3,580. | [APPENDIX] FFE regression of outcome on controls for child birth order, sex and age, and mother's birth year and age, mother FE, and indicator for HS participation. Robust standard errors clustered at mother's 1979 household. Specification does not mention inclusion of pre-treatment controls or indicator for other pre-k participation. Outcomes include high school graduation, some college, crime, teen parenthood and standardized (mean 0, std 1) index of adulthood outcomes. | [APPENDIX] No significant impact on high school graduation, some college, crime or teen parenthood. +0.26 *SD* (*SE* = 0.15) impact on index of adulthood outcomes. For African-Americans and "South" region, no significant impact on any outcome. For Male, 19.5% (*SE* = 0.10) effect on high school graduation, -22.0% (*SE* = 0.12) effect on crime and 0.54 *SD* (*SE* = 0.21) impact on index of adulthood outcomes. |

| REGRESSION DISCONTINUITY | | |
|---|---|---|
| Does Head Start Improve Children's Life Chances? Evidence from a Regression Discontinuity Design (Ludwig and Miller, 2007) Birth Cohorts: 1966-1978. Data includes OEO files from NARA of all federal expenditures for 1967-1980, child mortality from 1973-1983 Vital Statistics, county-level schooling from 1960-2000 decennial censuses, 1990 special tabulation from Census Bureau for schooling attainment by age, race and gender, restricted-use geo-coded NELS. | Non-parametric RD regression using local linear regressions with triangle kernel weights. RD cutoff based on which counties were eligible for assistance in applying for Head Start in 1965. Treatment group is the 300 poorest counties in 1965. Control group the next subsequent 300 poorest counties. Focus on average county level outcomes including child mortality rates, high school completion and some college. | **Overall:** HS participation increased high school completion rates (18-24, 1990) by 3-4% (p-values <0.05), increased some college (all cohorts) by around 3-5% (p-values < 0.05), and reduced childhood mortality (age 5-9) by 33-50% (p-values < 0.05). |

TABLE A1. (CONTINUED)

| | | |
|---|---|---|
| Long-Term Impacts of Compensatory Preschool on Health and Behavior: Evidence from Head Start (Carneiro and Ginja, 2014) Birth Cohorts: 1977-1996. Data includes NLSY mothers (born 1957-1965) and CNLSY children born to them. Focus on sample of children whose income between 15% - 185% for relevant income cutoff. Analysis Sample: N = 2,833. | Fuzzy RD regression using discontinuities in the probability of participation in HS created by eligibility rules. Outcomes include behaviors (e.g., drug use, obesity, grade repetition, alcohol use, school damage, smoking, drinking, sex, special education and BPI), cognitive (e.g., PIAT-M, PIAT-RR, PIAT-RC), crime (i.e., ever convicted or sentenced), CESD, high school diploma, birth control, idle, ever in college, ever work and standardized summary index composed of a weighted average of standardized (mean 0 and std 1) outcome variables. | **Reduced Form ITT** **Males ages 20-21:** HS has no significant impact on high school diploma, birth control, ever in college or ever worked. HS reduces the probability of crime by .40 ($SE = 20$), reduces idleness by 0.53 ($SE = 0.25$). Overall, no significant impact on summary index. **Males ages 16-17:** HS reduces probability of obesity by 0.47 ($SE = 0.19$), reduces CESD by 0.33 $SD$ ($SE = 0.10$). Overall, +0.16 $SD$ ($SE = 0.09$) impact on summary index. **Males ages 12-13:** HS reduces probability of obesity by 0.38 ($SE = 0.17$), reduces BPI by 0.27 $SD$ ($SE = 0.13$), reduces probability that health requires use of special equipment by 0.78 ($SE = 0.29$), reduces probability of need for frequent visits by doctors by 0.32 ($SE = 0.18$). No impact on test score measures. Overall, +0.21 $SD$ ($SE = 0.08$) impact on summary index. *Note:* Most effects found for children 12-13 years of age are driven by set of kids attending HS in the 1980s. **Structural Equations** **males ages 20-21:** HS participation had no significant impact on summary index or idleness. HS participation lead to a -22% reduction in crime **males ages 16-17:** HS participation had no significant impact on being overweight and summary index. -0.55 $SD$ in CESD **males ages 12-13:** HS participation lead to -29% in probability of being overweight, -29% in probability of needing special health equipment, +129% $SD$ in summary index (due to very fuzzy 1$^{st}$ stage). |

| DID + DID VARIANTS | | |
|---|---|---|
| Expanding Exposure: Can Increasing the Daily Exposure to Head Start Reduce Childhood Obesity? (Frisvold and Lumeng, 2011) Birth Cohorts: children of program years spanning 2001-2002 through 2005-2006. Data from 2001-2002 through 2005-2006 administrative data from a Michigan HS program. Sample: N = 1,833. | First model is DID regression of weight status at end of HS year on weight status at the beginning of the year, controls for individual and family characteristics, indicator for full-day HS participation, and year FE. Second model capitalizes on elimination of full-day HS expansion grants in 2003. 16 full-day classes in 2002 to only four in 2003. Implied IV specification where instrument is % of full-day funded slots. Controls include a battery of family characteristics. | **DID:** Full-day HS reduces obesity by 9.2 *pp* (*SE* = 0.026) (males), 6 *pp* (*SE* = 0.037) (black), no impact for females or white children. Overall, reduces obesity by 3.9 *pp* (*SE* = 0.018) (i.e., akin to change of 20 calories per day with no change in physical activity) **IV:** Full-day HS participation led to a decrease in obesity by 17.6 *pp*. F-stat on excluded instrument is 12.23. |
| Head Start's Long-Run Impact: Evidence from the Program's Introduction (Thompson, 2018) Birth Cohort: 1957-1964. Data from 1966–1970 CAP Records from NARA, 1968 and 1972 FOS files NARA (cross-validation), 1965, 1966 and 1968 OEO reports (cross-validation), NLSY respondents born 1957–1964, CNLSY respondents born to NLSY mothers. Analysis Sample: N = 2,685. Sample restricted to respondents who were ages 2-7 at the time of local Head Start implementation and who did not have a parent who attended college. | Regression of adult outcome on race, gender, birth order, number of siblings and maternal education controls, county of birth and cohort FE, and 3-year average of county-level of HS spending (birth cohort age 3-6). Outcomes: own income, household income, unemployment, educational attainment, high school dropout, college graduate, self-rate health, health conditions, health limitations and standardized adulthood summary index. Study design measures the effect of exposure to HS funding rather than direct impact of Head Start participation. | **Overall:** 0.125 (*SE* = 0.05) effect of HS exposure on educational attainment, 0.02 (*SE* = 0.01) effect on college graduation, $2,199 (*SE* = 877.15) effect on own income, $2,918 (*SE* = 1,437.38) effect on household income, and -0.05 (*SE* = 0.02) effect on health limitation, and 0.081 *SD* (*SE* = 0.02) on composite index of child's long-term outcomes, **Subgroup Analyses:** 0.06 *SD* (*SE* = 0.03) effect of HS exposure on composite index of child's long-term outcomes for whites, 0.09 *SD* (*SE* = 0.04) effect of HS exposure on composite index of child's long-term outcomes for blacks. |

*(continued)*

TABLE A1. (CONTINUED)

| | | |
|---|---|---|
| Breaking the Cycle? Intergenerational Effects of an Anti Poverty Program in Early Childhood (Barr and Gibbs, 2019) [Working Paper] Child Birth Cohorts: 2nd generation 1966–1968 CAP Records and 1968–1980 FOS Files from NARA for HS availability in FY 1966-1968, NLSY respondents born 1960–1964, CNLSY respondents born to NLSY mothers. Full sample: N = 3,533 (2nd gen child) and N = 2,398 (1st gen mother). High Impact sample: N = 1,687 (2nd gen child) and N = 821 (1st gen mother). Low Impact sample: N = 2,732 (2nd gen child) and N = 1,398 (1st gen mother). | Regression of adult outcome for a child on sex, age, age squared, race and mother's birth order controls, county of birth and birth year FE, and indicator variable for HS availability for a mother by birth cohort and county of birth. Outcomes: teen parenthood, interaction with criminal justice system (arrests, convictions or probations), high school graduation, some college attendance, and a adulthood summary index. Heterogeneity tests by race, gender and geographic location. Study design measures the effect of exposure to HS funding rather than direct impact of Head Start participation. | **High Impact Sample:** 0.67 ($SE = 0.23$) years of educational attainment, +12.7% ($SE = 0.05$) on high school graduation, 16.9% ($SE = 0.06$) effect on some college, -8.6% ($SE = 0.03$) effect on teen pregnancy, -15.6% ($SE = 0.04$) effect on crime, 0.47 $SD$ ($SE = 0.10$) effect on index of child's adulthood outcomes. Max educational attainment not included in main results as many individuals not yet finished with their education. **Low Impact Sample:** 0.28 ($SE = 0.18$) years of educational attainment, 6.4% ($SE = 0.04$) effect on high school graduation, 7.0% ($SE = 0.05$) effect on some college, -5.8% ($SE = 0.03$) effect on teen pregnancy, -6.3% ($SE = 0.03$) effect on crime, 0.22 ($SE = 0.09$) effect on index of child's adulthood outcomes. Max educational attainment not included in main results as many individuals not yet finished with their education. **Subgroup Analyses:** Larger effect on crime and smaller effects on teen parenthood for male children |
| Prep School for Poor Kids: The Long-Run Impacts of Head Start on Human Capital and Economic Self-Sufficiency (Bailey, Sun & Timpe, 2018) [Working Paper] Birth Cohorts: 1950-1980 Links long-form 2000 Census and 2001-2013 ACS with SSA's Numident file. Focus on individuals age 25-54 years old. | Event study framework where adult outcome of a child regressed on county-level controls, FEs for county of birth, year, and state-by-birth year, HS indicator interacted with child's school age at time of HS's launch. Outcomes: human capital index of high school or GED, some college, 4-year college degree, professional or doctoral degree, years of schooling, and indicator for occupation. Self-sufficiency index of employment, poverty status, various sources of income; continuous measures of weeks worked, usual hours worked, log ratio of family income to the federal poverty threshold. | **Overall:** HS caused +0.29 year (ATET 95% CI [.14, .49]) in schooling, +2.1% (ATET 95% CI [0.005, 0.038]) in high-school completion, +8.7% [ATET 95% CI [0.027, 0.092]) in college enrollment, +19% (ATET 95% CI [0.025, 0.094]) in college completion, and +.10 $SD$ on adult human capital index. HS caused about +.04 $SD$ on economic self-sufficiency index: 12% reduction in adult poverty; 29% in public assistance receipt. No statistically significant impacts of HS on incarceration. **Heterogeneity:** Differential impact of HS on human capital index by Medicaid exposure (ITT $F = 4.1$, $p<0.05$), CHC exposure (ITT $F = 9.3$, $p<0.01$), Food Stamps exposure (ITT $F = 4.5$, $p<0.05$), predicted economic growth (ITT $F = 11.9$, $p<0.01$). Differential impact of HS on self-sufficiency index by Food Stamps exposure (ITT $F = 3.5$, $p<0.06$). |

*(continued)*

TABLE A1. (CONTINUED)

| Study | Method | Results |
|---|---|---|
| Reducing Inequality Through Dynamic Complementarity: Evidence from Head Start and Public School Spending (Johnson and Jackson, 2019) Birth Cohorts: 1950-1976. Data includes annual HS spending at county level from NARA, ICPSR and SEER, annual public K12 spending at school district level, SFR database, 1968-2015 PSID data for individual long-term outcomes, 1960 Census data for county-level characteristics, and a multitude of datasets to capture info on timing of other key policy changes. | First model uses DID instrumental variables (DID-2SLS) where within-county, across-cohort DID variation in HS spending is exploited. Both public K12 spending and the interaction between HS spending per poor 4-year old and public K12 spending are instrumented using 2SLS. Second model is uses DID instrumental variables where all where all spending variables are instrumented (2SLS-2SLS). This includes HS spending, public K12 spending and their interaction. Outcomes include educational attainment, high school graduation, wages, incarceration and adult poverty. | **HS Spending (DID-2SLS):** For poor children, +0.08 ($SE = 0.02$) years of education, +2.5% ($SE = 0.007$) likelihood of high school graduation, 2.3% ($SE = 0.005$) higher wages (ages 20-50), +0.6% ($SE = 0.003$) less likelihood of adult incarceration, +1.8% ($SE = 0.005$) less likelihood of adult poverty. No significant impact for non-poor children. **Public K12 Spending (DID-2SLS):** Large and significant positive impacts for poor and non-poor children across probability of high school graduation, years of completed education, wages (age 20-50), less likelihood of adult incarceration and less likelihood of adult poverty. **Dynamic Complementarity (DID-2SLS):** For poor children exposed to typical HS center, a 10% increase in K12 spending leads to +0.59 ($SE = 0.12$) years of education, +14.8% ($SE = 0.02$) likelihood of high school graduation, +17.1% ($SE = 0.04$) higher wages (ages 20-50), +4.7% ($SE = 0.01$) less likelihood of adult incarceration, +12.2% ($SE = 0.04$) less likelihood of adult poverty |
| **INSTRUMENTAL VARIABLES** | | |
| Head Start Participation and Childhood Obesity (Frisvold, 2006) Birth Cohorts: ages 5-19 by 2002 Data from PSID, PSID CDS and U.S. Census Bureau's SAIPE. Sample: N = 2,301. Sub-Sample (white): N = 1,138. Sub-Sample (black): N = 973. | IV regression using a bivariate probit model where Head Start participation is instrumented by the number of spaces available in a community. Outcomes include childhood overweight and obesity (ages 5-19). | **Overall:** No significant impact of HS on obesity (ages 5-19). Significant impact of HS on overweight (ages 5-19) [ATT: -0.25, $SE = 0.08$] **Black:** Significant impact of HS on overweight (ages 5-19) [ATT: -0.334, $SE = 0.157$] and obesity (ages 5-19) [ATT: -0.332, $SE = 0.158$] **White:** No significant impact of HS on overweight and obesity (ages 5-19). |

ONLINE ONLY APPENDIX

The subsequent sections comprise a brief review of Head Start previous evaluations on short and medium term outcomes (**S1**), followed by a series of tables mentioned in the main text (**Tables S5-S20**) and in note 10 (**Tables S2-S4**).

**S1.** Review of Head Start Impacts on Short and Medium-Term Outcomes

What are the impacts of Head Start for children at the end or soon after program completion? Morris et al. (2018) found in a recent reanalysis of the Head Start Impact Study (HSIS; a 2002 national randomized evaluation of Head Start), results consistent with two other HSIS reanalyzes, similarly considering counterfactual types (Zhai, Brooks-Gunn, and Waldfogel 2014; Kline and Walters 2016). All three studies concluded that attending Head Start vs. no preschool yielded short benefits on cognitive test scores (measured 1 year after treatment) which quickly faded during the elementary school years. Morris et al., however, also found that centers' quality, geographic locations, counterfactual care alternatives, and dual language learner status, *all* moderated the treatment effect for some cognitive outcomes (e.g., receptive vocabulary; early numeracy). The authors thus emphasized interpreting estimated impacts in light of counterfactual and contextual condition comparisons. Finally, Kline and Walters projected impacts of test scores onto adulthood earnings gains and—after accounting for program substitution fiscal externalities—derived a greater than 1 benefit-cost ratio.

Accounting for previous Head Start evaluations' heterogeneous results, Shager et al.'s meta-analysis (2013) covering 28 Head Start studies, showed that about 41percent of cross-study differences were accounted by the quality of the evaluation research design. Lesser quality being mostly associated with *not* including an indicator distinguishing between 'active' (other preschool participation) and 'passive' (no preschool) control group. The authors summarized the effect sizes at 0.27 *SD* for cognitive and achievement outcomes measured less than a year after treatment. They also found that when the control group attended some of form of preschool, effect sizes on the same outcomes were smaller at 0.08 *SD* and not statistically significant.

Shager and colleagues (2013) did not specify for the kind of instruction delivered to the active control group. Instead, they considered whether outcomes of interest (i.e., skills) responded to any instruction. Camilli et al. (2010) conducted a meta-analysis spanning 128 center-based early intervention evaluations (about a fourth of which on Head Start programs) conducted from 1960 to 2000. They found that an individualized and explicit teacher-led form of instruction yielded an immediate boost on cognitive outcomes compared to a student-directed learning structure. Their overall preschool impact estimation of 0.23 *SD* on cognitive outcomes (versus no-preschool) was in line with Shager et al.'s results.

Equally consistent with other previous findings on the HSIS (e.g., Puma et al. 2012; Walters 2015), Camilli et al. (2010) found that early impacts tended to fade rapidly after intervention: this was especially the case when contrasting with a passive counterfactual group. Fadeout was less pronounced when comparing to a preschool alternative control group, which suggested that fadeout was proportionally related to corresponding impact magnitude. It was much lower in the case of preschool alternatives counterfactual. However, the authors could not reject the persistence into middle school (at or after age 10) of positive impacts on school and socio-emotional multiple outcomes. Deming (2009) found similar patterns of Head Start early cognitive boosts vanishing by age 11, along a persistent favorable impact (0.27 *SD*) on a non-test score composite of grade retention and learning disability diagnosis measured by age 14. Neither

Walters's nor Deming's studies reported significant impacts on psychometric measures of socio-emotional outcomes.

Investigating the relations between Head Start characteristics inputs and medium-term effectiveness (HSIS followed children up to third grade), Walters (2015) tested for interactions between seven center characteristic indicators and Head Start participation. Among these, only interactions with the variable 'delivery of a full day of service' (0.14 *SD*; *SE* = 0.06), and 'any staff having a teaching license' (0.13 *SD* = 0.07) were marginally impactful on cognitive outcomes. This last interaction, however, had a potentially moderate impact (up to 0.22 *SD*, within the 95 percent confidence interval), in a multivariate model including all interactions at once. However, estimates were not robust when applying the analysis to medium-run ones. For Walters, this indicated that notwithstanding program characteristics inputs' moderation of short-run impacts, the latter faded out all the same.

From one perspective, given the well-known impacts in adulthood of small RCTs of early childhood education programs targeted at economically disadvantaged children in the 1960s (see Elango et al., 2016 for review), the short-run benefits observed in the HSIS and Deming's sibling comparison study forecasted beneficial effects in adulthood for children who attended Head Start. In their cost-benefit analysis of Head Start, Kline and Walters (2016; and reference therein) noted that other educational interventions have yielded positive adult outcomes despite fadeout of test score gains. Deming's cost-benefit analysis projected earnings based on his estimates of Head Start effects on positive outcomes in early adulthood, which circumvents the problem of projecting long-term effects on the basis of short-term effects.

From a different perspective, the pattern of full fadeout of effects of Head Start on measures of academic achievement ran in notable contrast to findings from both the Perry

Preschool and Abecedarian programs, both of which produced persistent impacts on academic achievement into participants' 20s (Campbell et al. 2002; Schweinhart et al. 2005). Therefore, the underlying developmental processes may not have been causally impacted in the same way in recent Head Start cohorts as in these classic early childhood education studies, and therefore may yield different patterns of adult impacts.

References

Camilli, G., Vargas, S., Ryan, S., & Barnett, W. S. (2010). Meta-analysis of the effects of early education interventions on cognitive and social development. *Teachers college record*, *112*(3), 579-620.

Campbell, F. A., Ramey, C. T., Pungello, E., Sparling, J., & Miller-Johnson, S. (2002). Early childhood education: Young adult outcomes from the Abecedarian Project. *Applied developmental science*, *6*(1), 42-57.

Elango, S., García, J. L., Heckman, J. J., & Hojman, A. (2015). Early childhood education. In *Economics of Means-Tested Transfer Programs in the United States, Volume 2* (pp. 235-297). University of Chicago Press.

Kline, P., & Walters, C. R. (2016). Evaluating public programs with close substitutes: The case of Head Start. *The Quarterly Journal of Economics*, *131*(4), 1795-1848.

Morris, P. A., Connors, M., Friedman-Krauss, A., McCoy, D. C., Weiland, C., Feller, A., ... & Yoshikawa, H. (2018). New findings on impact variation from the Head Start Impact Study: Informing the scale-up of early childhood programs. *AERA Open*, *4*(2), n2.

Puma, M., Bell, S., Cook, R., Heid, C., Broene, P., Jenkins, F., ... & Downer, J. (2012). Third Grade Follow-Up to the Head Start Impact Study: Final Report. OPRE Report 2012-45. *Administration for Children & Families*.

Schweinhart, L. J., Montie, J., Xiang, Z., Barnett, S.W., Belfield, C. R. & Nores, M. (2005). Lifetime effects: The high/scope Perry preschool study through age 40 (monographs of the high/scope educational research foundation, 14). *Ypsilanti, MI: High Scope Educational Research Foundation*.

Shager, H. M., Schindler, H. S., Magnuson, K. A., Duncan, G. J., Yoshikawa, H., & Hart, C. M. (2013). Can research design explain variation in Head Start research results? A meta-analysis of cognitive and achievement outcomes. *Educational Evaluation and Policy Analysis*, *35*(1), 76-95.

Walters, C. R. (2015). Inputs in the production of early childhood human capital: Evidence from Head Start. *American Economic Journal: Applied Economics*, *7*(4), 76-102.

Zhai, F., Brooks-Gunn, J., & Waldfogel, J. (2014). Head Start's impact is contingent on alternative type of care in comparison group. *Developmental psychology*, *50*(12), 2572.

**Table S2.** Household Selected Characteristics by Preschool Status Across Cohorts

| | Head Start | | | | Preschool | | | | No preschool | | | | HS–None[c] |
|---|---|---|---|---|---|---|---|---|---|---|---|---|---|
| | Deming's cohort | Complement cohort[a] | Combined cohorts | p-value[b] | Deming's cohort | Complement cohort | Combined cohorts | p-value | Deming's cohort | Complement cohort | Combined cohorts | p-value | |
| *Permanent income* | 32,884 [21,810] | 39,800 [27,539] | 35,970 [24,830] | <.001 | 62,076 [41,520] | 92,664 [107,713] | 80,810 [89,545] | <.001 | 42,764 [30,000] | 61,857 [49,704] | 52,445 [41,989] | <.001 | -.39 |
| Fixed effects subsample | 34,672 [25,443] | 40,465 [29,118] | 37,571 [26,961] | .002 | 49,165 [29,600] | 77,243 [62,153] | 72,957 [55,368] | <.001 | 41,587 [27,968] | 61,793 [53,019] | 53,938 [45,831] | <.001 | -.36 |
| *Mother < high school* | .24 [.43] | .12 [.33] | .19 [.39] | <.001 | .08 [.27] | .04 [.20] | .06 [.23] | <.001 | .24 [.43] | .12 [.32] | .18 [.38] | <.001 | .03 |
| Fixed effects subsample | .28 [.45] | .14 [.35] | .20 [.40] | <.001 | .12 [.32] | .06 [.24] | .08 [.28] | .001 | .22 [.42] | .12 [.33] | .16 [.37] | <.001 | .11 |
| *Mother some college* | .28 [.45] | .40 [.49] | .33 [.47] | <.001 | .45 [.50] | .64 [.48] | .57 [.50] | <.001 | .25 [.43] | .44 [.50] | .34 [.48] | <.001 | -.02 |
| Fixed effects subsample | .25 [.43] | .42 [.49] | .34 [.47] | <.001 | .32 [.47] | .56 [.50] | .48 [.50] | <.001 | .27 [.44] | .43 [.50] | .37 [.48] | <.001 | -.06 |
| *Maternal AFQT* | -.61 [.61] | -.50 [.71] | -.56 [.66] | .001 | .01 [.85] | .49 [1.03] | .31 [1.00] | <.001 | -.36 [.80] | .11 [1.03] | -.12 [.95] | <.001 | -.46 |
| Fixed effects subsample | -.62 [.61] | -.48 [.70] | -.54 [.66] | .001 | -.21 [.81] | .19 [1.02] | .03 [.97] | <.001 | -.21 [.81] | -.01 [.99] | -.18 [.92] | <.001 | -.39 |
| *Grandmother's education* | 9.16 [3.09] | 9.69 [3.08] | 9.39 [3.09] | .001 | 10.69 [3.09] | 11.40 [3.06] | 11.13 [3.00] | <.001 | 9.45 [3.23] | 10.41 [3.39] | 9.92 [3.35] | <.001 | -.16 |
| Fixed effects subsample | 9.13 [3.12] | 9.79 [3.00] | 9.50 [3.04] | .001 | 10.10 [3.06] | 10.78 [3.22] | 10.51 [3.13] | <.001 | 9.69 [3.14] | 10.29 [3.30] | 10.07 [3.22] | <.001 | -.18 |
| Sample size | 779 | 637 | 1,491 | | 994 | 1,681 | 2,724 | | 1,931 | 1,857 | 3,658 | | |
| Sample size FE | 435 | 475 | 972 | | 459 | 835 | 1,349 | | 769 | 1,098 | 1,799 | | |

*Notes.* Standard deviations are in brackets. Permanent income is in 2014 dollars. [a] Complement cohort includes children deemed eligible from 1992 to 2000. [b] *p*-value corresponds to mean difference tests between Deming's cohort and Complement cohort, both included in the Combined cohorts. [c] HS-None: Combined cohorts mean difference (in *SD* units) between Head Start vs. No preschool.

**Table S3.** Household Selected Characteristics by Preschool Status Across Cohorts (White/Hispanic)

| | Head Start | | | | | Preschool | | | | | No preschool | | | | | |
|---|---|---|---|---|---|---|---|---|---|---|---|---|---|---|---|---|
| | Deming (2009)[A] | Deming's cohort | Complement cohort[a] | Combined cohorts | p-value[b] | Deming (2009) | Deming's cohort | Complement cohort | Combined cohorts | p-value | Deming (2009) | Deming's cohort | Complement cohort | Combined cohorts | p-value | HS–None[d] |
| *Permanent income*[c] | 26,553 [19,555] 34,770 [23,900] | 27,674 [19,022] | 42,645 [30,648] | 38,531 [27,449] | <.001 | 52,130 [34,577] 68,816 [43,233] | 54,771 [34,410] | 101,193 [116,519] | 88,968 [97,500] | <.001 | 35,592 [23,460] 46,298 [30,000] | 36,848 [23,877] | 66,956 [52,278] | 57,139 [44,583] | <.001 | -.42 |
| Fixed effects subsample | 27,560 [22,902] 35,783 [27,397] | 28,479 [21,805] | 45,261 [32,866] | 40,150 [30,177] | .001 | 41,882 [22,403] 54,926 [28,713] | 43,715 [22,852] | 87,466 [67,507] | 72,957 [55,368] | <.001 | 35,901 [23,600] 46,673 [29,614] | 37,150 [23,569] | 69,387 [57,066] | 61,637 [50,400] | <.001 | -.43 |
| *Mother < high school* | .51 [.50] | .35 [.48] | .16 [37] | .26 [.44] | <.001 | .18 [.38] | .08 [.26] | .04 [.20] | .05 [.23] | <.001 | .42 [.49] | .27 [.44] | .12 [.33] | .20 [.40] | <.001 | .15 |
| Fixed effects subsample | .53 [.50] | .34 [.47] | .18 [.39] | .27 [.44] | <.001 | .25 [.43] | .10 [.31] | .06 [.24] | .08 [.27] | .014 | .41 [.49] | .25 [.43] | .13 [.34] | .17 [.38] | <.001 | .26 |
| *Mother some college* | .22 [.41] | .22 [.41] | .32 [.47] | .26 [.44] | .004 | .41 [.49] | .42 [49] | .65 [.48] | .56 [.50] | <.001 | .23 [.42] | .23 [.42] | .44 [.50] | .34 [.47] | <.001 | -.17 |
| Fixed effects subsample | .16 [.37] | .16 [.36] | .33 [.47] | .24 [.43] | <.001 | .31 [.46] | .32 [.47] | .56 [.50] | .46 [.50] | <.001 | .22 [.41] | .23 [.43] | .41 [.49] | .35 [.48] | <.001 | -.23 |
| *Maternal AFQT* | -.44 [.73] | -.45 [.71] | -.31 [.80] | -.39 [.75] | .015 | .23 [.85] | .19 [.82] | .68 [1.00] | .50 [.97] | <.001 | -.21 [.86] | -.23 [.84] | .27 [1.05] | .03 [.98] | <.001 | -.43 |
| Fixed effects subsample | -.48 [.70] | -48 [.69] | -.27 [.77] | -.37 [.73] | .003 | .02 [.83] | -.01 [.80] | .42 [1.02] | .25 [.98] | <.001 | -.20 [.82] | -.22 [.80] | .18 [1.02] | .02 [.96] | <.001 | -.41 |
| *Grandmother's education* | 8.53 [3.50] | 8.53 [3.50] | 8.73 [3.57] | 8.63 [3.52] | .469 | 10.62 [2.92] | 10.62 [2.92] | 11.49 [3.18] | 11.17 [3.13] | <.001 | 9.34 [3.36] | 9.35 [3.36] | 10.36 [3.60] | 9.87 [3.52] | <.001 | -.35 |
| Fixed effects subsample | 8.51 [3.42] | 8.51 [3.42] | 8.96 [3.54] | 8.73 [3.45] | .175 | 10.09 [3.19] | 10.09 [3.19] | 10.81 [3.47] | 10.52 [3.36] | .002 | 9.54 [3.34] | 9.54 [3.34] | 10.25 [3.55] | 10.04 [3.44] | <.001 | -.38 |
| Sample size | 364 | 364 | 291 | 682 | | 745 | 745 | 1,359 | 2,148 | | 1,374 | 1,380 | 1,482 | 2,795 | | |
| Sample size FE | 229 | 229 | 211 | 478 | | 315 | 315 | 619 | 985 | | 510 | 510 | 827 | 1,269 | | |

*Notes.* Standard deviations are in brackets. [A] Deming's (2009) published estimates; [a] Complement cohort includes children deemed eligible from 1992 to 2000. [b] *p*-value corresponds to mean difference tests between Deming's cohort and complement cohort. [c] Permanent income in the Deming's cohort column is first in 2004 dollars; and below in 2014 dollars. [d] HS-None: Combined cohorts mean difference (in *SD* units) between Head Start vs. No preschool.

**Table S4.** Household Selected Characteristics by Preschool Status Across Cohorts (Black)

| | Head Start | | | | | Preschool | | | | | No preschool | | | | | |
|---|---|---|---|---|---|---|---|---|---|---|---|---|---|---|---|---|
| | Deming (2009)[A] | Deming's cohort | Complement cohort[a] | Combined cohorts | p-value[b] | Deming (2009) | Deming's cohort | Complement cohort | Combined cohorts | p-value | Deming (2009) | Deming's cohort | Complement cohort | Combined cohorts | p-value | HS–None[d] |
| *Permanent income*[c] | 24,005 [16,103] | 24,856 [15,661] 31,231 [19,678] | 37,406 [24,412] | 33,810 [22,174] | <.001 | 32,470 [21,939] | 33,353 [18,639] 41,907 [27,398] | 56,668 [40,940] | 50,387 [36,069] | <.001 | 25,980 [18,496] | 26,993 [18,639] 33,914 [23,419] | 41,704 [30,449] | 37,243 [27,097] | <.001 | -.13 |
| Fixed effects subsample | 26,010 [19,559] | 26,612 [18,370] 33,436 [23,081] | 36,632 [25,147] | 35,076 [27,097] | .157 | 28,940 [22,853] | 29,102 [21,983] 36,564 [27,620] | 47,947 [26,878] | 43,238 [27,269] | <.001 | 24,164 [16,314] | 25,128 [16,815] 31,572 [21,128] | 38,615 [27,247] | 35,504 [23,829] | .001 | -.02 |
| *Mother < high school* | .33 [.47] | .15 [.36] | .09 [.29] | .13 [.33] | .009 | .20 [.40] | .08 [.27] | .05 [.22] | .06 [.23] | .136 | .38 [.49] | .17 [.37] | .09 [.28] | .13 [.34] | <.001 | .00 |
| Fixed effects subsample | .39 [.49] | .21 [.41] | .11 [.31] | .14 [.35] | .003 | .27 [.45] | .14 [.35] | .07 [.26] | .09 [.29] | .045 | .37 [.48] | .19 [.39] | .08 [.27] | .14 [.34] | <.001 | .00 |
| *Mother some college* | .31 [.46] | .34 [.48] | .47 [.50] | .40 [.49] | .001 | .50 [.50] | .53 [.50] | .60 [.49] | .58 [.49] | .102 | .28 [.45] | .30 [.46] | .47 [.50] | .37 [.48] | <.001 | .06 |
| Fixed effects subsample | .32 [.47] | .35 [.48] | .49 [.50] | .44 [.50] | .002 | .42 [.50] | .48 [.50] | .56 [.50] | .53 [.50] | .111 | .30 [.46] | .35 [.48] | .49 [.50] | .41 [.49] | .001 | .06 |
| *Maternal AFQT* | -.75 [.49] | -.75 [.47] | -.66 [.57] | -.71 [.53] | .004 | -.51 [.72] | -.52 [.71] | -.33 [.76] | -.41 [.74] | .003 | -.68 [.60] | -.68 [.59] | -.54 [.65] | -.62 [.63] | .001 | -.14 |
| Fixed effects subsample | -.77 [.48] | -.78 [.47] | -.64 [.59] | -.69 [.55] | | -.63 [.66] | -.63 [.65] | -.45 [.66] | -.56 [.63] | .011 | -.76 [.56] | -.77 [.54] | -.56 [.65] | -.66 [.59] | <.001 | -.05 |
| *Grandmother's education* | 9.71 [2.56] | 9.71 [2.56] | 10.51 [2.30] | 10.02 [2.50] | <.001 | 10.88 [2.68] | 10.88 [2.68] | 10.98 [2.44] | 10.98 [2.49] | .636 | 9.70 [2.87] | 9.70 [2.87] | 10.61 [2.40] | 10.09 [2.72] | <.001 | -.03 |
| Fixed effects subsample | 9.82 [2.59] | 9.82 [2.59] | 10.44 [2.27] | 10.25 [2.34] | .006 | 10.13 [2.76] | 10.13 [2.76] | 10.71 [2.36] | 10.48 [2.40] | .035 | 9.98 [2.67] | 9.98 [2.67] | 10.43 [2.42] | 10.14 [2.61] | .042 | .04 |
| Sample size | 415 | 415 | 346 | 809 | | 249 | 249 | 322 | 576 | | 551 | 551 | 375 | 863 | | |
| Sample size FE | 206 | 206 | 264 | 494 | | 144 | 144 | 216 | 364 | | 259 | 259 | 271 | 530 | | |

*Notes.* Standard deviations are in brackets. [A] Deming's (2009) published estimates; [a] Complement cohort includes children deemed eligible from 1992 to 2000. [b] *p*-value corresponds to mean difference tests between Deming's cohort and complement cohort. [c] Permanent income in the Deming's cohort column is first in 2004 dollars; and below in 2014 dollars. [d] HS-None: Combined cohorts mean difference (in *SD* units) between Head Start vs. No preschool.

**Table S5.** Outcomes means [& standard deviations] by preschool status across cohorts

| | Head Start | | | | Preschool | | | | No preschool | | | | |
|---|---|---|---|---|---|---|---|---|---|---|---|---|---|
| | Deming's cohort | Complement cohort | Combined cohorts | p-value[a] | Deming's cohort | Complement cohort | Combined cohorts | p-value | Deming's cohort | Complement cohort | Combined cohorts | p-value | HS - No p-value[b] |
| *Adult summary index (SD)* | -.29 [1.06] | -.08 [.96] | -.21 [1.02] | .00 | -.06 [1.01] | .35 [.82] | .19 [.92] | <.00 | -.29 [1.06] | .22 [.90] | -.03 [1.02] | <.00 | .97/ <.00/ <.00 |
| *HS graduate* | .71 [.46] | .81 [.40] | .75 [.43] | .00 | .78 [.41] | .89 [.32] | .85 [.36] | <.00 | .74 [.44] | .85 [.36] | .79 [.41] | <.00 | .25/.04/ .03 |
| *Some college* | .39 [.49] | .45 [.50] | .41 [.49] | .06 | .53 [.50] | .62 [.49] | .59 [.49] | .01 | .40 [.49] | .55 [.50] | .48 [.50] | <.00 | .72/ <.00/ <.00 |
| *Idle* | .22 [.41] | .22 [.41] | .21 [.41] | .88 | .15 [.36] | .12 [.32] | .13 [.34] | .14 | .22 [.41] | .13 [.33] | .17 [.38] | <.00 | .95/ <.00/ .01 |
| *Crime* | .36 [.48] | .29 [.45] | .35 [.48] | .03 | .36 [.48] | .19 [.39] | .26 [.44] | <.00 | .38 [.49] | .21 [.41] | .30 [.46] | <.00 | .48/.00/ .02 |
| *Teen pregnancy* | .33 [.47] | .20 [.40] | .25 [.44] | <.00 | .24 [.43] | .11 [.31] | .15 [.36] | <.00 | .29 [.45] | .12 [.32] | .20 [.40] | <.00 | .27/ <.00/ < .00 |
| *Poor health status* | .11 [.32] | .16 [.37] | .16 [.36] | .06 | .16 [.36] | .09 [.29] | .11 [.32] | .00 | .17 [.37] | .13 [.38] | .14 [.35] | .09 | .05/ .09/ .28 |
| *Cognitive tests index (SD)* | -.52 [.84] | -.42 [.82] | -.47 [.83] | .08 | -.18 [.92] | .12 [.96] | -.02 [.96] | <.00 | -.39 [.93] | -.05 [.99] | -.24 [.98] | <.00 | .04/ <.00/ <.00 |
| *Nontests index (SD)* | -.10 [.96] | -.12 [1.18] | -.14 [1.11] | .76 | -.02 [1.04] | .15 [.89] | .09 [.95] | .01 | -.20 [1.04] | .17 [.87] | .01 [.96] | <.00 | .14/ <.00/ <.00 |
| *Learning disability* | .04 [.19] | .09 [.28] | .07 [.25] | .00 | .06 [.23] | .05 [.21] | .05 [.22] | .48 | .05 [.22] | .04 [.20] | .05 [.21] | .39 | .31/ <.00/ .01 |
| *Grade retention* | .37 [.48] | .27 [.45] | .33 [.47] | .00 | .27 [.45] | .18 [.38] | .21 [.41] | <.00 | .42 [.49] | .18 [.38] | .28 [.45] | <.00 | .21/ <.00/ .01 |
| Sample size | 350 | 464 | 951 | | 346 | 739 | 1275 | | 495 | 747 | 1512 | | |

*Notes*. Standard deviations are in brackets. [a] *p*-values correspond to mean difference tests between Deming's cohort and Complement cohort, both included in the Combined cohorts. [b] HS – No *p*-values correspond to mean difference tests between Head Start vs. No preschool status for Deming's/Complement/Combined cohorts, in that order. Adulthood summary index (standardized) and its 6 composing individual outcomes were measured up to the 2014 CNLSY survey-round. Cognitive test index (standardized) is a composite of the average of 3 standardized test scores: PPVT, PIAT Math, and PIAT Reading Recognition. Nontests index (standardized) comprises two indicators: grade retention and learning disability status. Positive index values signal more desirable outcomes and negative index values signal less desirable outcomes.

**Figure S6.** Distributions of outcome indexes by preschool status across cohorts

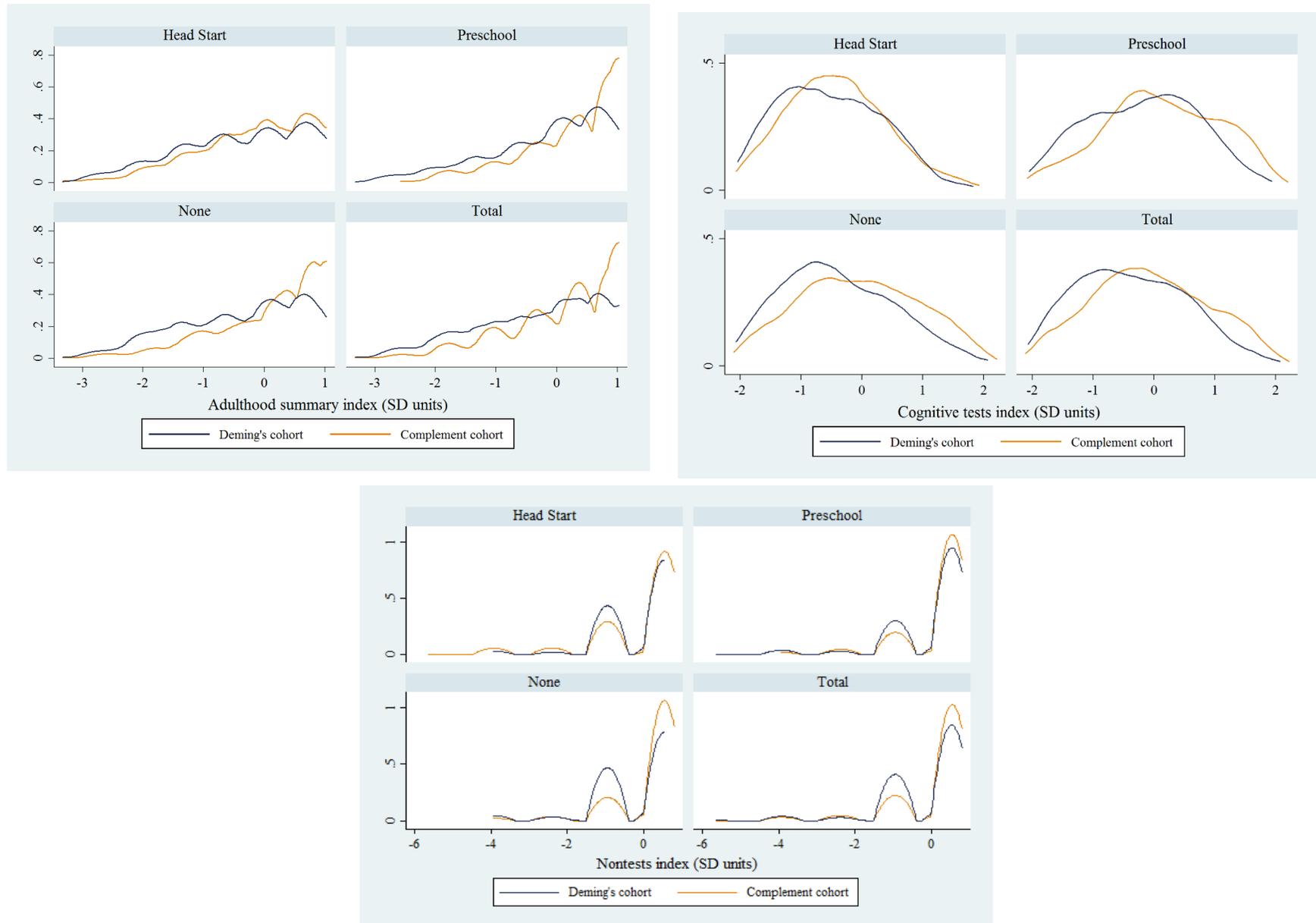

*Notes.* Distributions were smoothed, and densities estimated via the Epanechnikov kernel function

**Table S7.** Sibling Differences on Pre-Treatment Covariates by Preschool Status

| | Head Start | | Preschool | | No preschool (Control mean) | |
|---|---|---|---|---|---|---|
| | Complement cohort | Combined cohorts | Complement cohort | Combined cohorts | Complement cohort | Combined cohorts |
| Pre-treatment index | -0.001 | 0.005 | 0.031 | 0.018 | -0.003 | -0.031 |
| | (0.063) | (0.035) | (0.039) | (0.027) | [0.975]/2,144 | [0.983]/3,738 |
| Attrition | **-0.029** | -0.009 | **-0.023** | -0.011 | 0.045 | 0.025 |
| | (0.013) | (0.007) | (0.012) | (0.007) | [0.207]/2,215 | [0.169]/3,826 |
| Age in 2014 (in years) | **1.07** | 0.085 | **1.80** | **0.415** | 24.45 | 29.03 |
| | (0.425) | (0.272) | (0.270) | (0.212) | [3.95]/2,144 | [5.75]/3,738 |
| PPVT at age 3 | 1.58 | 1.78 | -0.535 | -1.01 | 22.00 | 21.05 |
| | (3.38) | (1.99) | (3.73) | (2.05) | [13.25]/396 | [12.60]/644 |
| Log of birth weight | -0.008 | 0.023 | **-0.025** | -0.010 | 4.76 | 4.73 |
| | (0.018) | (0.012) | (0.012) | (0.009) | [0.203]/1,894 | [0.225]/3,449 |
| Very low BW | 0.006 | -0.007 | 0.010 | 0.005 | 0.006 | 0.013 |
| | (0.010) | (0.007) | (0.007) | (0.004) | [0.074]/1,894 | [0.114]/3,449 |
| In mother's HH, age 0-3 | 0.003 | 0.008 | -0.002 | -0.001 | 0.988 | 0.974 |
| | (0.007) | (0.007) | (0.005) | (0.005) | [0.085]/2,123 | [0.119]/3,601 |
| Previous health limitation | 0.001 | 0.008 | **-0.045** | -0.020 | 0.095 | 0.066 |
| | (0.026) | (0.014) | (0.018) | (0.012) | [0.293]/2,123 | [0.249]/3,601 |
| Firstborn | 0.100 | -0.016 | **0.277** | 0.051 | 0.183 | 0.319 |
| | (0.052) | (0.031) | (0.039) | (0.026) | [0.387]/2,144 | [0.466]/3,738 |
| Male | -0.019 | -0.004 | -0.054 | -0.010 | 0.528 | 0.512 |
| | (0.043) | (0.026) | (0.115) | (0.023) | [0.499]/2,144 | [0.500]/3,738 |
| HOME score at age 3 | 0.829 | -0.247 | **4.49** | **2.84** | 42.48 | 42.00 |
| | (2.35) | (1.75) | (1.67) | (1.34) | [26.16]/1,649 | [26.64]/2,106 |
| Father in HH, age 0-3 | 0.006 | -0.000 | -0.018 | 0.002 | 0.759 | 0.714 |
| | (0.027) | (0.021) | (0.016) | (0.014) | [0.402]/2,033 | [0.427]/2,891 |
| Grandmother in HH, age 0-3 | 0.010 | -0.018 | 0.021 | -0.014 | 0.110 | 0.175 |
| | (0.661) | (0.016) | (0.138) | (0.010) | [0.267]/2,137 | [0.316]/3,622 |
| Maternal care, age 0-3 | **-0.059** | -0.023 | **-0.089** | **-0.052** | 0.618 | 0.650 |
| | (0.026) | (0.016) | (0.020) | (0.013) | [0.413]/2,114 | [0.410]/3,702 |
| Relative care, age 0-3 | 0.008 | 0.013 | 0.026 | 0.021 | 0.184 | 0.177 |
| | (0.023) | (0.014) | (0.018) | (0.012) | [0.320]/2,114 | [0.323]/3,702 |
| Non-relative care, age 0-3 | **0.051** | 0.010 | **0.063** | **0.031** | 0.198 | 0.173 |
| | (0.025) | (0.013) | (0.018) | (0.012) | [0.336]/2,114 | [0.320]/3,702 |
| Breastfed | 0.053 | -0.011 | -0.008 | 0.002 | 0.484 | 0.412 |
| | (0.029) | (0.017) | (0.022) | (0.014) | [0.500]/1,933 | [0.492]/3,511 |
| Regular doctor visits, age 0-3 | -0.012 | -0.018 | 0.012 | -0.015 | 0.476 | 0.468 |
| | (0.053) | (0.040) | (0.035) | (0.029) | [0.500]/1,918 | [0.499]/2,360 |
| Ever at dentist, age 0-3 | -0.043 | -0.037 | -0.035 | -0.016 | 0.227 | 0.228 |
| | (0.049) | (0.037) | (0.039) | (0.032) | [0.419]/1,449 | [0.420]/1,798 |
| Weight change during pregnancy | -1.08 | -0.138 | -0.238 | -0.780 | 31.24 | 30.66 |
| | (1.17) | (0.692) | (0.857) | (.549) | [14.96]/1,873 | [14.98]/3,337 |
| Illness, age 0-1 | 0.025 | 0.004 | 0.013 | -0.014 | 0.564 | 0.537 |
| | (0.043) | (0.026) | (0.033) | (0.021) | [0.496]/1,907 | [0.499]/3,401 |
| Premature birth | 0.031 | -0.016 | -0.033 | -0.014 | 0.231 | 0.228 |
| | (0.037) | (0.021) | (0.028) | (0.018) | [0.422]/1,902 | [0.416]/3,400 |
| Private health insurance, age 0-3 | 0.004 | -0.015 | -0.019 | -0.014 | 0.697 | 0.655 |
| | (0.033) | (0.026) | (0.018) | (0.017) | [0.422]/1,925 | [0.446]/2,368 |
| Medicaid, age 0-3 | -0.038 | 0.008 | -0.028 | -0.015 | 0.265 | 0.267 |
| | (0.026) | (0.024) | (0.015) | (0.014) | [0.411]/1,925 | [0.417]/2,366 |

**Table S7.** Sibling Differences on Pre-Treatment Covariates by Preschool Status

|  | Head Start | | Preschool | | No preschool (Control mean) | |
| --- | --- | --- | --- | --- | --- | --- |
|  | Complement cohort | Combined cohorts | Complement cohort | Combined cohorts | Complement cohort | Combined cohorts |
| Log income, age 0-3 | 0.020 | 0.012 | -0.051 | -0.014 | 10.62 | 10.46 |
|  | (0.048) | (0.030) | (0.033) | (0.022) | [0.935]/2,052 | [0.848]/3,546 |
| Log income at age 3 | 0.044 | 0.018 | 0.036 | 0.019 | 10.59 | 10.43 |
|  | (0.068) | (0.045) | (0.050) | (0.034) | [1.05]/1,734 | [0.971]/2,952 |
| Mom avg. hours worked, year before birth | -2.24 | -1.89 | 0.239 | -0.004 | 30.23 | 28.79 |
|  | (1.35) | (1.12) | (0.970) | (0.773) | [14.10]/1,564 | [13.54]/1,905 |
| Mom avg. hours worked, age 0-1 | -1.15 | -0.453 | 0.763 | 1.30 | 34.05 | 32.95 |
|  | (1.23) | (1.16) | (0.963) | (0.752) | [12.77]/1,144 | [12.26]/1,639 |
| Mom smoked before birth | 0.005 | 0.001 | 0.030 | 0.016 | 0.294 | 0.336 |
|  | (0.024) | (0.017) | (0.018) | (0.013) | [0.456]/1,917 | [0.472]/3,430 |
| Mom drank before birth | 0.003 | 0.001 | 0.010 | 0.003 | 0.059 | 0.083 |
|  | (0.022) | (0.013) | (0.494) | (0.010) | [0.235]/2,144 | [0.276]/3,738 |

*Notes*: Standard errors are in parenthesis. Estimates in bold were significant at the 5 percent level or less. Standard deviations are in brackets, followed by cohort sample size for that covariate. Each covariate was regressed on preschool status indicators (Head Start; other 'Preschool') with 'No preschool' as the reference status (i.e., the control mean).

**Table S8.** Selection Bias Estimates: Sibling Differences in Pre-Treatment Covariates by Preschool Status Across Cohorts

| | Head Start estimate (SE) | | | | Preschool estimate (SE)/p-value[b] | | | | None: Control mean [SD]/Overall sample size | | |
|---|---|---|---|---|---|---|---|---|---|---|---|
| | Deming's cohort | Complement cohort | Combined cohorts | p-value[a] | Deming's cohort | Complement cohort | Combined cohorts | p-value[a] | Deming's cohort | Complement cohort | Combined cohorts |
| Pre-treatment index | .014 (.061) | -.001 (.063) | .005 (.035) | .098 | .047 (.055) | .031 (.039) | .018 (.027) | .536 | -.063 [.986]/1251 | -.003 [.975]/2,144 | -.031 [.983]/3,738 |
| PPVT at 3 | 2.24 (4.82) | 1.58 (3.38) | 1.78 (1.99) | .774 | -7.16* (4.12)/.070 | -.535 (3.73) | -1.01 (2.05) | .740 | 19.90 [11.10]/195 | 22.00 [13.25]/396 | 21.05 [12.60]/644 |
| Log of birth weight | .048** (.020) | -.008 (.018) | .023* (.012) | .031 | -.006 (.017)/.064 | -.025** (.012) | -.010 (.009)/.015 | .506 | 4.70 [.25]/1,226 | 4.76 [.203]/1,894 | 4.73 [.225]/3,449 |
| Very low BW | -.022* (.012) | .006 (.010) | -.007 (.007) | .095 | -.004 (.008) | .010 (.007) | .005 (.004) | .227 | .021 [.145]/1,226 | .006 [.074]/1,894 | .013 [.114]/3,449 |
| In mother's HH, 0-3 | .002 (.029) | .003 (.007) | .008 (.007) | .807 | -.028 (.027) | -.002 (.005) | -.001 (.005) | .952 | .900 [.302]/1,187 | .988 [.085]/2,123 | .974 [.119]/3,601 |
| Previous health limitation | -.001 (.014) | .001 (.026) | .008 (.014) | .387 | -.041** (.018)/.052 | -.045** (.018)/.078 | -.020* (.012)/.064 | .750 | .405 [.197]/1,187 | .095 [.293]/2,123 | .066 [.249]/3,601 |
| Firstborn | .016 (.055) | .100* (.052) | -.016 (.031) | .024 | -.124** (.055)/.038 | .277*** (.039)/.0009 | .051* (.026)/.046 | <.001 | .419 [.494]/1,251 | .183 [.387]/2,144 | .319 [.466]/3,738 |
| Male | -.000 (.048) | -.019 (.043) | -.004 (.026) | .927 | -.003 (.046) | -.054 (.115) | -.010 (.023) | .092 | .501 [.500]/1,251 | .528 [.499]/2,144 | .512 [.500]/3,738 |
| HOME at 3 | 1.98 (3.25) | .829 (2.35) | -.247 (1.75) | .693 | 3.03 (4.10) | 4.49** (1.67) | 2.84** (1.34)/.087 | .350 | 38.05 [26.25]/427 | 42.48 [26.16]/1,649 | 42.00 [26.64]/2,106 |
| Father in HH, 0-3 | .009 (.034) | .006 (.027) | -.000 (.021) | .580 | -.003 (.023) | -.018 (.016) | .002 (.014) | .747 | .624 [.450]/739 | .759 [.402]/2,033 | .714 [.427]/2,891 |
| Grandmother in HH, 0-3 | -.003 (.024) | .010 (.661) | -.018 (.016) | .627 | -.049*** (.019)/.078 | .021 (.138) | -.014 (.010) | .024 | .215 [.325]/1,190 | .110 [.267]/2,137 | .175 [.316]/3,622 |
| Maternal care, 0-3 | .0187 (.019) | -.059** (.026) | -.023 (.016) | .064 | -.015 (.483) | -.089*** (.020) | -.052*** (.013)/.084 | .097 | .689 [.405]/1,244 | .618 [.413]/2,114 | .650 [.410]/3,702 |
| Relative care, 0-3 | -.007 (.019) | .008 (.023) | .013 (.014) | .232 | .022 (.019) | .026 (.018) | .021* (.012) | .766 | .180 [.335]/1,244 | .184 [.320]/2,114 | .177 [.323]/3,702 |
| Non-relative care, 0-3 | -.012 (.017) | .051** (.025) | .010 (.013) | .387 | -.006 (.016) | .063*** (.018) | .031** (.012) | .020 | .131 [.283]/1,244 | .198 [.336]/2,114 | .173 [.320]/3,702 |
| Breastfed | -.053** (.027) | .053* (.029) | -.011 (.017) | .093 | -.010 (.024) | -.008 (.022)/.042 | .002 (.014) | .994 | .333 [.472]/1,234 | .484 [.500]/1,933 | .412 [.492]/3,511 |

**Table S8.** Selection Bias Estimates: Sibling Differences in Pre-Treatment Covariates by Preschool Status Across Cohorts

| | Head Start estimate (SE) | | | | Preschool estimate (SE)/p-value[b] | | | | None: Control mean [SD]/Overall sample size | | |
|---|---|---|---|---|---|---|---|---|---|---|---|
| | Deming's cohort | Complement cohort | Combined cohorts | p-value[a] | Deming's cohort | Complement cohort | Combined cohorts | p-value[a] | Deming's cohort | Complement cohort | Combined cohorts |
| Regular doctor visits, 0-3 | .043 (.102) | -.012 (.053) | -.018 (.040) | .865 | -.055 (.110) | .012 (.035) | -.015 (.029) | .707 | .383 [.488]/430 | .476 [.500]/1,918 | .468 [.499]/2,360 |
| Ever at dentist, 0-3 | .033 (.137) | -.043 (.049) | -.037 (.037) | .586 | .008 (.137) | -.035 (.039) | -.016 (.032) | .583 | .303 [.461]/401 | .227 [.419]/1,449 | .228 [.420]/1,798 |
| Weight change during pregnancy | .056 (1.18) | -1.08 (1.17) | -.138 (.692) | .840 | -.168 (1.14) | -.238 (.857) | -.780 (.549) | .692 | 29.71 [15.34]/1,146 | 31.24 [14.96]/1,873 | 30.66 [14.98]/3,337 |
| Illness, 0-1 | .016 (.042) | .025 (.043) | .004 (.026) | .721 | -.061 (.041) | .013 (.033) | -.014 (.021) | .123 | .520 [.500]/1,175 | .564 [.496]/1,907 | .537 [.499]/3,401 |
| Premature birth | -.047 (.034) | .031 (.037) | -.016 (.021) | .093 | .007 (.034) | -.033 (.028) | -.014 (.018) | .255 | .218 [.413]/1,175 | .231 [.422]/1,902 | .228 [.416]/3,400 |
| Private health insurance, 0-3 | .093 (.069) | .004 (.033) | -.015 (.026) | .096 | .032 (.049) | -.019 (.018) | -.014 (.017) | .151 | .447 [.481]/431 | .697 [.422]/1,925 | .655 [.446]/2,368 |
| Medicaid, 0-3 | .048 (.061) | -.038 (.026) | .008 (.024) | .984 | -.006 (.043) | -.028* (.015) | -.015 (.014) | .526 | .376 [.456]/431 | .265 [.411]/1,925 | .267 [.417]/2,366 |
| Log income, 0-3 | -.012 (.040) | .020 (.048) | .012 (.030) | .589 | .040 (.033) | -.051 (.033) | -.014 (.022) | .624 | 10.00 [.718]/1,186 | 10.62 [.935]/2,052 | 10.46 [.848]/3,546 |
| Log income, at 3 | .011 (.085) | .044 (.068) | .018 (.045) | .721 | .054 (.064) | .036 (.050) | .019 (.034) | .730 | 9.98 [.826]/993 | 10.59 [1.05]/1,734 | 10.43 [.971]/2,952 |
| Mom avg. hours worked, year b. birth | -1.11 (3.14) | -2.24* (1.35) | -1.89* (1.12) | .500 | 2.06 (1.87) | .239 (.970) | -.004 (.773)/.093 | .185 | 26.03 [12.15]/377 | 30.23 [14.10]/1,564 | 28.79 [13.54]/1,905 |
| Mom avg. hours worked, 0-1 | -1.08 (3.17) | -1.15 (1.23) | -.453 (1.16) | .797 | 1.77 (1.72) | .763 (.963) | 1.30* (.752) | .356 | 32.52 [11.07]/379 | 34.05 [12.77]/1,144 | 32.95 [12.26]/1,639 |
| Mom smoked b. birth | -.012 (.031) | .005 (.024) | .001 (.017) | .821 | -.005 (.023) | .030* (.018) | .016 (.013) | .399 | .392 [.489]/1,186 | .294 [.456]/1,917 | .336 [.472]/3,430 |
| Mom drank b. birth | .004 (.021) | .003 (.022) | .001 (.013) | .577 | .010 (.021) | .010 (.494) | .003 (.010) | .911 | .081 [.272]/1,251 | .059 [.235]/2,144 | .083 [.276]/3,738 |

*Notes.* *** $p < .01$; ** $p < .05$; * $p < .1$. [a] $p$-value corresponds to estimates difference tests between Deming's cohort and complement cohort, both included in the Combined cohorts. [b] $p$-value <.1: estimate difference test between Head Start and Preschool status.

**Table S9.** Selection Bias Estimates: Sibling Differences in Attrition and Age by Preschool Status Across Cohorts

| | Head Start estimate (SE) | Preschool estimate (SE) | None: Control mean [SD]/Overall sample size |
|---|---|---|---|
| Deming's cohort — Non-attrited[a] observations | .014 (.021) | .032 (.022) | .860 [.347]/1,384 |
| Complement cohort — Attrited observations | -.029** (.013) | -.023** (.012) | .045 [.207]/2,215 |
| Combined cohorts — Attrited observation | -.009 (.007) | -.011 (.007) | .025 [.169]/3,826 |
| Deming's cohort — Age in 2004[b] | .182 (.298) | -.433 (.249) | 23.20 [2.88]/1,251 |
| Complement cohort — Age in 2014 | 1.07** (.425) | 1.80*** (.270) | 24.45 [3.95]/2,144 |
| Combined cohorts — Age in 2014 | .085 (.272) | .415** (.212) | 29.03 [5.75]/3,738 |

*Notes.* *** $p < .01$; ** $p < .05$; * $p < .1$. [a] For the Deming's cohort, non-attrited observations count is considered instead of attrited due to different cross-round year of last interview variable available in NLSY, and used to derived attrition counts. Instead, it was possible to derive non-attrition count from by-survey-round dichotomous interview variables. It is non-attrition status that is used in FE sample eligibility: in Deming's cohort, the same non-attrition counts as in Deming (2009) was obtained. [b] Age by 2004 (i.e., at least 4 by 1990) in Deming's cohort is of course different when taken for the same cohort a decade later in the combined cohorts (i.e., at least 4 by 2000): therefore, these selection bias estimates are not readily comparable across cohorts unlike the other covariates and are considered separately.

**Table S10.** Head Start Impacts on Averaged Test & BPI Scores; Nontest & Adulthood Summary Index Scores, Across Cohorts (Overall & by Subgroups)

| | Test (5-14) | | | | | BPI (5-14) | | | | Nontest score index (7-14) | | | | | Adulthood index (19+) | | | | | |
|---|---|---|---|---|---|---|---|---|---|---|---|---|---|---|---|---|---|---|---|---|
| | Deming (2009)[a] | Repl.[b] | Comp.[c] | Comb.[d] | p[e] | Repl. | Comp. | Comb. | p | Deming (2009) | Repl. | Comp. | Comb. | p | Deming (2009) | Repl. | Ext.[f] (28+) | Comp. | Comb. | p |
| *Overall* | | | | | | | | | | | | | | | | | | | | |
| Head Start | .101 (.057) | .106* (.056) | -.003 (.062) | -.007 (.036) | .243 | .056 (.047) | .069 (.048) | .071** (.029) | .816 | .265*** (.082) | .262*** (.081) | -.149* (.080) | -.001 (.047) | .001 | .228*** (.072) | .202*** (.072) | .166** (.069) | -.145** (.068) | -.011 (.041) | .005 |
| Preschool | -.012 (.062) | .010 (.060) | .015 (.042) | .006 (.030) | .682 | .040 (.047) | .109*** (.040) | .074*** (.026) | .234 | .172* (.088) | .177** (.088) | -.124** (.060) | -.028 (.040) | .008 | .069 (.072) | .091 (.074) | .034 (.066) | -.040 (.053) | -.003 (.035) | .644 |
| p (HS=preschool) | .118 | .175 | .801 | .731 | | .762 | .369 | .924 | | .372 | .420 | .747 | .605 | | .080 | .237 | .134 | .148 | .859 | |
| Sample size [HS/Preschool] | 1,251 [364/364] | 1,251 [364/364] | 2,144 [497/795] | 3,738 [951/1,275] | | 1,251 [364/364] | 2,144 [497/795] | 3,738 [951/1,275] | | 1,251 [364/364] | 1,251 [364/364] | 2,141 [497/795] | 3,734 [951/1,275] | | 1,251 [364/364] | 1,251 [364/364] | 1,251 [364/364] | 2,144 [497/795] | 3,738 [951,1,275] | |
| *By race* | | | | | | | | | | | | | | | | | | | | |
| Head Start (white/Hispanic) | .110 (.090) | .110 (.090) | -.097 (.091) | -.054 (.054) | .081 | .094 (.063) | .098 (.068) | .112*** (.040) | .762 | .177 (.111) | .173 (.110) | -.155 (.108) | -.005 (.068) | .001 | .237** (.103) | .177 (.100) | .153 (.102) | -.087 (.102) | -.011 (.062) | .154 |
| Head Start (black) | .107 (.072) | .108 (.072) | .062 (.083) | .041 (.049) | .798 | .028 (.067) | .003 (.067) | .031 (.042) | .978 | .351*** (.120) | .348*** (.119) | -.135 (.122) | .008 (.066) | .036 | .224** (.102) | .234** (.106) | .182** (.092) | -.180* (.099) | -.017 (.056) | .012 |
| p (nonblack=black) | .982 | .988 | .191 | .184 | | .476 | .323 | .162 | | .282 | .277 | .900 | .896 | | .924 | .697 | .831 | .531 | .945 | |
| Sample size [nonblack/black] | 1,251 [695/556] | 1,251 [695/556] | 2,144 [1,375/769] | 3,738 [2,280/1,458] | | 1,251 [695/556] | 2,144 [1,375/769] | 3,738 [2,280/1,458] | | 1,251 [695/556] | 1,251 [695/556] | 2,141 [1,374/767] | 3,734 [2,278/1,456] | | 1,251 [695/556] | 1,251 [695/556] | 1,251 [695/556] | 2,144 [1,375/769] | 3,738 [2,280/1,458] | |
| *By gender* | | | | | | | | | | | | | | | | | | | | |
| Head Start (male) | .159** (.076) | .160** (.076) | -.065 (.085) | -.019 (.049) | .118 | .159** (.067) | .099 (.064) | .098** (.040) | .953 | .390*** (.123) | .385*** (.122) | -.223* (.117) | -.003 (.069) | .005 | .182* (.103) | .130 (.098) | .103 (.101) | -.195** (.100) | -.063 (.058) | .058 |
| Head Start (female) | .055 (.081) | .055 (.081) | .056 (.077) | .005 (.048) | .742 | -.035 (.064) | .037 (.067) | .044 (.041) | .805 | .146 (.108) | .144 (.108) | -.054 (.102) | .002 (.063) | .008 | .272** (.106) | .272*** (.96) | .226** (.112) | -.090 (.093) | .042 (.061) | .019 |
| p (male=female) | .346 | .343 | .246 | .716 | | .034 | .489 | .331 | | .135 | .140 | .262 | .961 | | .533 | .273 | .450 | .414 | .220 | |
| Sample size [male/female] | 1,251 [627/624] | 1,251 [627/624] | 2,144 [1,104/1,040] | 3,738 [1,904/1,834] | | 1,251 [627/624] | 2,144 [1,104/1,040] | 3,738 [1,904/1,834] | | 1,251 [627/624] | 1,251 [627/624] | 2,141 [1,102/1,039] | 3,734 [1,901/1,833] | | 1,251 [627/624] | 1,251 [627/624] | 1,251 [627/624] | 2,144 [1,104/1,040] | 3,738 [1,904/1,834] | |
| *By maternal AFQT score* | | | | | | | | | | | | | | | | | | | | |
| Head Start (AFQT ≤ -1) | .015 (.094) | -.006 (.094) | .054 (.136) | -.030 (.067) | .083 | -.065 (.062) | .037 (.101) | .017 (.050) | .946 | .529*** (.156) | .509*** (.153) | -.212 (.193) | .083 (.093) | .106 | .279** (.114) | .317*** (.119) | .384*** (.115) | .027 (.145) | .108 (.080) | .147 |
| Head Start (AFQT > -1) | .154** (.071) | .168** (.071) | -.022 (.068) | .004 (.043) | .008 | .124* (.064) | .079 (.053) | .091** (.036) | .802 | .124 (.091) | .123 (.091) | -.094 (.087) | -.028 (.054) | .003 | .202** (.091) | .166* (.090) | .046 (.084) | -.178** (.079) | -.053 (.048) | .041 |
| p (low=high AFQT) | .245 | .146 | .617 | .671 | | .034 | .709 | .227 | | .024 | .030 | .573 | .303 | | .595 | .309 | .017 | .218 | .086 | |
| Sample size [low/high] | 1,251 [365/886] | 1,251 [365/886] | 2,144 [365/1,779] | 3,738 [810/2,928] | | 1,251 [365/886] | 2,144 [365/1,779] | 3,738 [810/2,928] | | 1,251 [365/886] | 1,251 [365/886] | 2,141 [364/1,777] | 3,734 [808/2,926] | | 1,251 [365/886] | 1,251 [365/886] | 1,251 [365/886] | 2,144 [365/1,779] | 3,738 [810/2,928] | |

*Notes.* *** $p < .01$; ** $p < .05$; * $p < .1$. [a] Deming's (2009) published estimates. [b] Repl. = Deming (2009) replicated. [c] Comp. = Complement cohort includes siblings fitting the same criteria as in Deming (2009) but found eligible from 1990 to 2000. [d] Comb. = Combined cohorts includes siblings up to 2000 (i.e., integrate both Deming's and the complement cohorts). [e] *p*-value for estimates' difference testing between Ext. and Complement cohort. [f] Ext. = Deming's (2009) cohort (i.e., siblings eligible up to 1990), with outcomes extended to last available NLSY 2014 survey round (by then, siblings are all 28+ years old).

**Table S11.** Head Start Impacts on Individual Outcomes, Across Cohorts (Overall & by Subgroups)

| | Grade retention | | | | Learning disability diagnosis | | | | Teen parenthood | | | | High school graduation | | | | |
|---|---|---|---|---|---|---|---|---|---|---|---|---|---|---|---|---|---|
| | Repl.[a] | Comp.[b] | Comb.[c] | $p$ [d] | Repl. | Comp. | Comb. | $p$ | Repl. | Comp. | Comb. | $p$ | Repl. | Ext.[e] (28+) | Comp. | Comb. | $p$ |
| *Overall* | | | | | | | | | | | | | | | | | |
| Head Start | -.069* | .010 | -.005 | .137 | -.059*** | .042** | .003 | .002 | -.002 | .034 | -.002 | .338 | .087*** | .020 | -.000 | .007 | .905 |
| | (.040) | (.034) | (.022) | | (.021) | (.021) | (.012) | | (.036) | (.032) | (.021) | | (.032) | (.032) | (.029) | (.019) | |
| Preschool | -.086** | .022 | .006 | .003 | -.023 | .027* | .009 | .286 | -.057 | .027 | -.006 | .330 | -.006 | -.025 | .015 | -.005 | .218 |
| | (.037) | (.023) | (.018) | | (.031) | (.016) | (.010) | | (.035) | (.020) | (.016) | | (.030) | (.032) | (.020) | (.016) | |
| $p$ (HS=preschool) | .680 | .709 | .603 | | .187 | .480 | .673 | | .229 | .816 | .853 | | .015 | .262 | .599 | .549 | |
| *By race* | | | | | | | | | | | | | | | | | |
| Head Start (white/Hispanic) | -.029 | .029 | -.012 | .495 | -.046 | .034 | .010 | .002 | .014 | .030 | -.010 | .927 | .046 | .032 | -.054 | -.006 | .578 |
| | (.059) | (.044) | (.031) | | (.030) | (.029) | (.018) | | (.053) | (.041) | (.028) | | (.050) | (.050) | (.043) | (.029) | |
| Head Start (black) | -.103* | -.015 | -.002 | .143 | -.071** | .056* | -.004 | .134 | -.025 | .061 | .020 | .164 | .121*** | .007 | .066* | .018 | .434 |
| | (.055) | (.052) | (.033) | | (.028) | (.031) | (.015) | | (.051) | (.050) | (.030) | | (.041) | (.040) | (.040) | (.027) | |
| $p$ (nonblack=black) | .357 | .521 | .816 | | .527 | .594 | .547 | | .593 | .621 | .463 | | .255 | .695 | .042 | .535 | |
| *By gender* | | | | | | | | | | | | | | | | | |
| Head Start (male) | -.203*** | .056 | -.031 | .028 | -.047 | .044 | .013 | .043 | .032 | .025 | -.004 | .104 | .119** | .036 | -.029 | .001 | .799 |
| | (.058) | (.046) | (.033) | | (.030) | (.030) | (.017) | | (.053) | (.038) | (.026) | | (.050) | (.050) | (.042) | (.029) | |
| Head Start (female) | .056 | -.037 | .020 | .817 | -.070*** | .040* | -.007 | .002 | -.034 | .041 | .000 | .919 | .056 | .004 | .027 | .014 | .643 |
| | (.056) | (.044) | (.030) | | (.026) | (.023) | (.015) | | (.056) | (.048) | (.031) | | (.044) | (.044) | (.038) | (.026) | |
| $p$ (male=female) | .002 | .124 | .243 | | .563 | .913 | .374 | | .415 | .793 | .922 | | .368 | .646 | .323 | .745 | |
| *By maternal AFQT score* | | | | | | | | | | | | | | | | | |
| Head Start (AFQT ≤ -1) | -.128* | -.037 | -.027 | .833 | -108*** | .074 | -.018 | .099 | -.033 | -.023 | -.023 | .560 | .163*** | .059 | .188*** | .077* | .181 |
| | (.067) | (.076) | (.043) | | (.041) | (.048) | (.023) | | (.064) | (.075) | (.041) | | (.057) | (.053) | (.071) | (.039) | |
| Head Start (AFQT > -1) | -.033 | .013 | .001 | .143 | -.031 | .031 | .011 | .008 | .016 | .050 | .008 | .529 | .044 | -.004 | -.044 | -.017 | .847 |
| | (.050) | (.037) | (.026) | | (.022) | (.022) | (.013) | | (.043) | (.035) | (.023) | | (.037) | (.039) | (.031) | (.022) | |
| $p$ (low=high AFQT) | .252 | .543 | .586 | | .095 | .402 | .266 | | .530 | .370 | .508 | | .080 | .322 | .003 | .037 | |

*Notes.* *** $p < .01$; ** $p < .05$; * $p < .1$. [a] Repl. = Deming (2009) replicated. [b] Complement cohort includes siblings fitting the same criteria as in Deming (2009) but found eligible from 1990 to 2000. [c] Combined cohorts includes siblings up to 2000 (i.e., integrate both Deming's and the complement cohorts). [d] $p$-value for estimates' difference testing between Ext. and Complement cohort. [e] Ext. = Deming's (2009) cohort (i.e., siblings eligible up to 1990), with outcomes extended to last available NLSY 2014 survey round (by then, siblings are all 28+ years old).

**Table S12.** Head Start Impacts on Individual Outcomes, Across Cohorts (Overall & by Subgroups)

| | Some college attended | | | | | Idle | | | | | Crime | | | | | Poor health status | | | | |
|---|---|---|---|---|---|---|---|---|---|---|---|---|---|---|---|---|---|---|---|---|
| | Repl.[a] | Ext.[b] (28+) | Comp.[c] | Comb.[d] | $p$ [e] | Repl. | Ext. (28+) | Comp. | Comb. | $p$ | Repl. | Ext. (28+) | Comp. | Comb. | $p$ | Repl. | Ext. (28+) | Comp. | Comb. | $p$ |
| *Overall* | | | | | | | | | | | | | | | | | | | | |
| Head Start | .022 | .110*** | -.074* | -.002 | .010 | -.059* | .002 | .082*** | .016 | .026 | -.002 | -.028 | .033 | .009 | .071 | -069** | -.047 | -.025 | -.003 | .459 |
| | (.035) | (.039) | (.039) | (.023) | | (.035) | (.033) | (.030) | (.020) | | (.038) | (.041) | (.031) | (.021) | | (.027) | (.032) | (.031) | (.019) | |
| Preschool | .061* | .053 | -.043 | .010 | .030 | .005 | .019 | .028 | .005 | .458 | .001 | -.001 | .014 | .012 | .905 | -.020 | -.009 | -.033 | -.003 | .253 |
| | (.036) | (.042) | (.030) | (.020) | | (.029) | (.031) | (.023) | (.016) | | (.037) | (.036) | (.026) | (.018) | | (.032) | (.031) | (.021) | (.015) | |
| $p$ (HS=preschool) | .379 | .269 | .440 | .627 | | .114 | .648 | .074 | .566 | | .933 | .571 | .569 | .892 | | .153 | .327 | .796 | .993 | |
| *By race* | | | | | | | | | | | | | | | | | | | | |
| Head Start | -.046 | .108 | -.057 | -.011 | .071 | -.091* | -.012 | .070 | .027 | .047 | -.018 | -.044 | -.016 | -.007 | .417 | -092** | -.038 | -.074 | -.010 | .315 |
| (white/Hispanic) | (.048) | (.054) | (.053) | (.031) | | (.053) | (.047) | (.045) | (.028) | | (.058) | (.063) | (.043) | (.032) | | (.043) | (.054) | (.045) | (.029) | |
| Head Start | .083* | .118** | -.093* | .004 | .037 | -.036 | .014 | .079* | .000 | .242 | .010 | -.014 | .057 | .015 | .086 | -.047 | -.049 | .020 | .009 | .040 |
| (black) | (.050) | (.055) | (.057) | (.034) | | (.045) | (.049) | (.041) | (.028) | | (.050) | (.052) | (.044) | (.029) | | (.035) | (.037) | (.045) | (.025) | |
| $p$ (nonblack=black) | .061 | .891 | .643 | .741 | | .424 | .707 | .873 | .502 | | .710 | .704 | .240 | .601 | | .425 | .865 | .155 | .610 | |
| *By gender* | | | | | | | | | | | | | | | | | | | | |
| Head Start | -.029 | .075 | -.130** | -.024 | .063 | -.088* | -.025 | .111*** | .055** | .162 | .025 | -.043 | .101** | .042 | .331 | -.035 | -.020 | -107*** | -.020 | .612 |
| (male) | (.046) | (.054) | (.054) | (.031) | | (.045) | (.046) | (.042) | (.025) | | (.058) | (.060) | (.045) | (.031) | | (.037) | (.046) | (.038) | (.023) | |
| Head Start (female) | .070 | .144*** | -.016 | .019 | .028 | -.032 | .028 | .049 | -.023 | .056 | -.029 | -.014 | -.037 | -.024 | .073 | -103** | -.072 | .057 | .014 | .109 |
| | (.050) | (.053) | (.054) | (.033) | | (.048) | (.052) | (.044) | (.030) | | (.053) | (.057) | (.043) | (.029) | | (.042) | (.049) | (.042) | (.028) | |
| $p$ (male=female) | .135 | .350 | .130 | .342 | | .376 | .464 | .317 | .045 | | .505 | .740 | .027 | .128 | | .247 | .461 | .002 | .329 | |
| *By maternal AFQT score* | | | | | | | | | | | | | | | | | | | | |
| Head Start | .001 | .150*** | -.089 | .022 | .161 | -.064 | -.078 | .153** | .039 | .035 | -.009 | .011 | .050 | .024 | .466 | -091** | -157*** | -.074 | -078** | .299 |
| (AFQT ≤ -1) | (.049) | (.056) | (.078) | (.040) | | (.064) | (.062) | (.064) | (.041) | | (.068) | (.072) | (.062) | (.041) | | (.046) | (.059) | (.067) | (.037) | |
| Head Start | .034 | .094* | -.066 | -.009 | .025 | -.057 | .045 | .065* | .008 | .142 | .001 | -.051 | .011 | .000 | .107 | -.058* | .014 | -.007 | .027 | .903 |
| (AFQT > -1) | (.047) | (.052) | (.045) | (.028) | | (.042) | (.039) | (.033) | (.022) | | (.045) | (.049) | (.036) | (.025) | | (.034) | (.037) | (.035) | (.022) | |
| $p$ (low=high AFQT) | .621 | .473 | .800 | .529 | | .924 | .092 | .222 | .508 | | .905 | .477 | .587 | .626 | | .572 | .015 | .371 | .014 | |

*Notes.* *** $p < .01$; ** $p < .05$; * $p < .1$. [a] Repl. = Deming (2009) replicated. [b] Ext. = Deming's (2009) cohort (i.e., siblings eligible up to 1990), with outcomes extended to last available NLSY 2014 survey round (by then, siblings are all 28+ years old). [c] Comp. = Complement cohort includes siblings fitting the same criteria as in Deming (2009) but found eligible from 1990 to 2000. [d] Comb. = Combined cohorts includes siblings up to 2000 (i.e., integrate both Deming's and the complement cohorts). [e] $p$-value for estimates' difference testing between Deming's cohort and Complement cohort.

**Table S13.** Head Start/Preschool Longer-Run Impacts on Adulthood Index Score, Educational Outcomes & Earnings (Overall and by Subgroups)

|  | Adulthood index | | Educational attainment | College graduation | Earnings |
|---|---|---|---|---|---|
|  | Deming (2009) | Deming's cohort (extended outcomes)[a] | Deming's cohort (extended outcomes) | Deming's cohort (extended outcomes) | Deming's cohort (extended outcomes) |
| *Overall* | | | | | |
| Head Start | .228*** | .166** | .306** | -.016 | .067 |
|  | (.072) | (.069) | (.145) | (.028) | (.122) |
| Preschool | .069 | .034 | .138 | .006 | -.049 |
|  | (.072) | (.066) | (.169) | (.028) | (.114) |
| *p* (HS=preschool) | .080 | .134 | .410 | .452 | .429 |
| Sample size | 1,251 | 1,251 | 1,251 | 1,251 | 1,195 |
| [HS/Preschool] | [364/364] | [364/364] | [364/364] | [364/364] | [350/343] |
| *By race* | | | | | |
| Head Start | .224** | .153 | .269 | -.051* | .191 |
| (white/Hispanic) | (.102) | (.102) | (.203) | (.029) | (.193) |
| Head Start | .237** | .182** | .334 | .003 | -.037 |
| (black) | (.103) | (.092) | (.207) | (.032) | (.165) |
| *p* (nonblack=black) | .924 | .831 | .822 | .218 | .373 |
| Sample size | 1,251 | 1,251 | 1,251 | 1,251 | 1,195 |
| [nonblack/black] | [695/556] | [695/556] | [695/556] | [695/556] | [656/539] |
| *By gender* | | | | | |
| Head Start | .182* | .103 | .268 | -.004 | -.026 |
| (male) | (.103) | (.101) | (.206) | (.032) | (.174) |
| Head Start | .272** | .226** | .343* | -.028 | .153 |
| (female) | (.106) | (.112) | (.205) | (.033) | (.186) |
| *p* (male=female) | .553 | .450 | .797 | .627 | .501 |
| Sample size | 1,251 | 1,251 | 1,251 | 1,251 | 1,195 |
| [male/female] | [627/624] | [627/624] | [627/624] | [627/624] | [581/614] |
| *By maternal AFQT score* | | | | | |
| Head Start | .279** | .384*** | .454** | -.005 | .437** |
| (AFQT ≤ -1) | (.114) | (.115) | (.227) | (.031) | (.197) |
| Head Start | .202** | .046 | .238 | -.022 | -.132 |
| (AFQT > -1) | (.091) | (.084) | (.187) | (.029) | (.153) |
| *p* (low=high AFQT) | .595 | .017 | .462 | .688 | .023 |
| Sample size | 1,251 | 1,251 | 1,251 | 1,251 | 1,195 |
| [low/high] | [365/886] | [365/886] | [365/886] | [365/886] | [354/841] |

*Notes.* [a] Deming's cohort (i.e., siblings eligible up to 1990), with outcomes extended to last available NLSY 2014 survey round (by then, siblings are all 28+ years old). Adulthood index (in standard deviation) = composite of 6 indicators for high school graduation; college attendance; teen-age parenthood; 'idle'; involvement with the justice system; and poor health status. Educational attainment = amount of years of completed schooling. College graduation indicator (1= 16+ years of completed schooling); Earnings = natural log of averaged yearly earnings, in 2014 dollars, adjusted for age and survey-round year. Pretreatment covariates and sibling fixed effects are included throughout. 'No preschool' status is the counterfactual. Standard errors, in parenthesis, are clustered at the family level. *** $p < .01$; ** $p < .05$; * $p < .1$.

**Table S14.** Head Start & Preschool Impacts on Cognitive Test Scores Across Cohorts (by age groups)

| | (1) | | | | | (2) | | | | | (3) | | | | | (4) | | | | | (5) | | | | |
|---|---|---|---|---|---|---|---|---|---|---|---|---|---|---|---|---|---|---|---|---|---|---|---|---|---|
| | Deming (2009)[a] | Rep.[b] | Comp.[c] | Comb.[d] | p[e] | Deming (2009) | Rep. | Comp. | Comb. | p | Deming (2009) | Rep. | Comp. | Comb. | p | Deming (2009) | Rep. | Comp. | Comb. | p | Deming (2009) | Rep. | Comp. | Comb. | p |
| **Head Start** | | | | | | | | | | | | | | | | | | | | | | | | | |
| 5-6 | -.025 (.091) | -.027 (.091) | -.249*** (.065) | -.159*** (.052) | .074 | .081 (.083) | .080 (.083) | -.016 (.063) | .043 (.049) | .536 | .093 (.079) | .103 (.079) | -.001 (.061) | .054 (.048) | .470 | .131 (.087) | .129 (.087) | .055 (.070) | .053 (.048) | .122 | .145* (.085) | .145* (.085) | .062 (.069) | .041 (.048) | .239 |
| 7-10 | -.116 (.072) | -.117 (.072) | -.334*** (.059) | -.233*** (.046) | .038 | .040 (.065) | .039 (.065) | -.108* (.057) | -.026 (.042) | .299 | .067 (.061) | .078 (.060) | -.090* (.053) | -.008 (.039) | .185 | .116* (.060) | .116* (.060) | -.037 (.063) | -.005 (.039) | .067 | .133** (.060) | .133** (.060) | -.030 (.062) | -.013 (.040) | .115 |
| 11-14 | -.201*** (.070) | -.201*** (.070) | -.319*** (.063) | -.233*** (.045) | .088 | -.053 (.065) | -.053 (.065) | -.101* (.059) | -.039 (.041) | .486 | -.017 (.061) | -.009 (.060) | -.083 (.056) | -.020 (.039) | .395 | .029 (.061) | .029 (.061) | -.053 (.067) | -.028 (.040) | .231 | .055 (.062) | .056 (.062) | -.048 (.066) | -.035 (.040) | .255 |
| **Preschool** | | | | | | | | | | | | | | | | | | | | | | | | | |
| 5-6 | .167** (.083) | .161* (.083) | .264*** (.052) | .233*** (.043) | .211 | .022 (.082) | .014 (.081) | .112** (.049) | .085** (.040) | .442 | -.019 (.078) | -.024 (.077) | .094* (.048) | .053 (.039) | .285 | -.102 (.084) | -.107 (.084) | .095* (.051) | .031 (.038) | .098 | -.079 (.085) | -.081 (.085) | .079 (.051) | .017 (.039) | .109 |
| 7-10 | .230*** (.070) | .228*** (.70) | .232*** (.049) | .237*** (.037) | .955 | .111* (.064) | .108* (.064) | .067 (.044) | .094*** (.033) | .243 | .087 (.061) | .087 (.061) | .053 (.042) | .068** (.032) | .290 | .031 (.061) | .031 (.061) | .030 (.045) | .034 (.031) | .960 | .048 (.065) | .049 (.065) | .015 (.046) | .024 (.032) | .875 |
| 11-14 | .182** (.072) | .182** (.072) | .204*** (.055) | .189*** (.040) | .768 | .076 (.068) | .074 (.068) | .032 (.048) | .056 (.036) | .133 | .037 (.065) | .039 (.064) | .004 (.046) | .019 (.034) | .148 | -.040 (.066) | -.039 (.066) | -.032 (.049) | -.015 (.035) | .865 | -.022 (.069) | -.020 (.069) | -.040 (.049) | -.023 (.035) | .856 |
| Permanent income | | | | | | | | | | | .112* (.064) | .118** (.055) | .102*** (.033) | .100*** (.028) | | | | | | | | | | | |
| Maternal AFQT | | | | | | | | | | | .353*** (.057) | .345*** (.057) | .259*** (.028) | .241*** (.026) | | | | | | | | | | | |
| Mom HS graduate | | | | | | | | | | | .141** (.071) | .238*** (.074) | .241*** (.065) | .272*** (.050) | | | | | | | | | | | |
| Mom some coll. | | | | | | | | | | | .280*** (.080) | .395*** (.088) | .318*** (.072) | .411*** (.055) | | | | | | | | | | | |
| p all age Head Start effects = | .074 | .077 | .323 | .231 | | .096 | .100 | .269 | .226 | | .161 | .151 | .279 | .290 | | .092 | .097 | .187 | .217 | | .151 | .155 | .185 | .257 | |
| Baseline covariates | N | N | N | N | | Y | Y | Y | Y | | Y | Y | Y | Y | | N | N | N | N | | Y | Y | Y | Y | |
| Sibling FE | N | N | N | N | | N | N | N | N | | N | N | N | N | | Y | Y | Y | Y | | Y | Y | Y | Y | |
| Total # of scores | 4,687 | 4,687 | 8,220 | 14,086 | | 4,687 | 4,687 | 8,220 | 14,086 | | 4,687 | 4,687 | 8,220 | 14,086 | | 4,687 | 4,687 | 8,220 | 14,086 | | 4,687 | 4,687 | 8,220 | 14,086 | |
| $R^2$ | .028 | .028 | .056 | .053 | | .194 | .195 | .232 | .236 | | .268 | .273 | .293 | .295 | | .608 | .607 | .633 | .574 | | .619 | .618 | .644 | .582 | |
| Sample size | 1,251 | 1,251 | 2,144 | 3,738 | | 1,251 | 1,251 | 2,144 | 3,738 | | 1,251 | 1,251 | 2,144 | 3,738 | | 1,251 | 1,251 | 2,144 | 3,738 | | 1,251 | 1,251 | 2,144 | 3,738 | |

*Notes:* Test scores are standardized (mean = 0; SD = 1). *** $p < .01$; ** $p < .05$; * $p < .1$. [a] Deming's (2009) published estimates. [b] Repl. = Deming (2009) replicated. [c] Comp. = Complement cohort includes siblings fitting the same criteria as in Deming (2009) but found eligible from 1990 to 2000. [d] Comb. = Combined cohorts includes siblings up to 2000 (i.e., integrate both Deming's and the complement cohorts). [e] $p$-value for estimates' difference testing between Deming (2009) replicated. and Complement cohort. For further details on models (1) to (5) specifications refer to Deming (2009).

**Table S15.** Head Start & Preschool Impacts on BPI Scores Across Cohorts (by age groups)

| | (1) | | | | (2) | | | | (3) | | | | (4) | | | | (5) | | | |
|---|---|---|---|---|---|---|---|---|---|---|---|---|---|---|---|---|---|---|---|---|
| | Repl.[a] | Comp.[b] | Comb.[c] | $p$[d] | Repl. | Comp. | Comb. | $p$ | Repl. | Comp. | Comb. | $p$ | Repl. | Comp. | Comb. | $p$ | Repl. | Comp. | Comb. | $p$ |
| Head Start | | | | | | | | | | | | | | | | | | | | |
| 5-6 | .122 | .74 | .140*** | .587 | .097 | -.036 | .058 | .520 | .080 | -.022 | .063 | .552 | .046 | .024 | .030 | .906 | .063 | .060 | .031 | .878 |
| | (.094) | (.063) | (.052) | | (.314) | (.063) | (.051) | | (.094) | (.063) | (.050) | | (.085) | (.062) | (.045) | | (.083) | (.061) | (.045) | |
| 7-10 | .149** | .134** | .194*** | .509 | .102 | .017 | .106** | .831 | .084 | .030 | .108*** | .679 | .044 | .079 | .085** | .215 | .053 | .064 | .085** | .274 |
| | (.066) | (.055) | (.042) | | (.067) | (.055) | (.041) | | (.066) | (.055) | (.040) | | (.057) | (.050) | (.033) | | (.055) | (.050) | (.033) | |
| 11-14 | .098 | .126** | .139*** | .870 | .062 | .025 | .061 | .337 | .042 | .038 | .064 | .435 | .054 | .067 | .049 | .764 | .063 | .052 | .048 | .500 |
| | (.068) | (.061) | (.044) | | (.068) | (.061) | (.043) | | (.067) | (.060) | (.043) | | (.063) | (.058) | (.038) | | (.063) | (.058) | (.038) | |
| Preschool | | | | | | | | | | | | | | | | | | | | |
| 5-6 | .074 | -.103** | -.060 | .023 | .113 | -.078 | -.038 | .089 | .117 | -.065 | -.022 | .068 | .047 | .053 | .043 | .951 | .082 | .052 | .045 | .762 |
| | (.083) | (.052) | (.042) | | (.082) | (.050) | (.042) | | (.082) | (.051) | (.042) | | (.073) | (.048) | (.037) | | (.071) | (.050) | (.038) | |
| 7-10 | .099 | -.054 | -.021 | .141 | .117 | -.030 | .000 | .259 | .123 | -.024 | .012 | .220 | .054 | .090** | .075** | .310 | .077 | .089** | .076** | .284 |
| | (.064) | (.046) | (.035) | | (.064) | (.045) | (.035) | | (.063) | (.045) | (.035) | | (.055) | (.042) | (.030) | | (.055) | (.044) | (.031) | |
| 11-14 | -.044 | .002 | -.013 | .923 | -.022 | .040 | .005 | .821 | -.020 | .044 | .019 | .782 | -.003 | .132*** | .078** | .229 | .023 | .131*** | .074** | .321 |
| | (.067) | (.049) | (.037) | | (.066) | (.048) | (.037) | | (.066) | (.048) | (.037) | | (.057) | (.046) | (.031) | | (.058) | (.048) | (.032) | |
| Permanent income | | | | | | | | | | -.175*** | -.089*** | -.133*** | | | | | | | | |
| | | | | | | | | | (.061) | (.034) | (.031) | | | | | | | | | |
| Maternal AFQT | | | | | | | | | | .082 | .080*** | .086*** | | | | | | | | |
| | | | | | | | | | (.055) | (.030) | (.027) | | | | | | | | | |
| Mom HS graduate | | | | | | | | | | -.242*** | -.153* | -.238*** | | | | | | | | |
| | | | | | | | | | (.087) | (.096) | (.061) | | | | | | | | | |
| Mom some coll. | | | | | | | | | | -.294*** | -.221** | -.307*** | | | | | | | | |
| | | | | | | | | | (.096) | (.088) | (.067) | | | | | | | | | |
| $p$ all age Head Start effects = | .77 | .540 | .264 | | .849 | .572 | .390 | | .839 | .583 | .410 | | .990 | .536 | .360 | | .987 | .514 | .350 | |
| Baseline covariates | N | N | N | | Y | Y | Y | | Y | Y | Y | | N | N | N | | Y | Y | Y | |
| Sibling FE | N | N | N | | N | N | N | | N | N | N | | Y | Y | Y | | Y | Y | Y | |
| Total # of scores | 4,610 | 8,303 | 14,098 | | 4,610 | 8,303 | 14,098 | | 4,610 | 8,303 | 14,098 | | 4,610 | 8,303 | 14,098 | | 4,610 | 8,303 | 14,098 | |
| $R^2$ | .025 | .049 | .060 | | .065 | .108 | .101 | | .080 | .115 | .114 | | .540 | .579 | .532 | | .553 | .584 | .534 | |
| Sample size | 1,251 | 2,144 | 3,738 | | 1,251 | 2,144 | 3,738 | | 1,251 | 2,144 | 3,738 | | 1,251 | 2,144 | 3,738 | | 1,251 | 2,144 | 3,738 | |

*Notes:* BPI scores are standardized (mean = 0; SD = 1). *** $p < .01$; ** $p < .05$; * $p < .1$. [a] Repl. = Deming (2009) replicated. [b] Comp. = Complement cohort includes siblings fitting the same criteria as in Deming (2009) but found eligible from 1990 to 2000. [c] Comb. = Combined cohorts includes siblings up to 2000 (i.e., integrate both Deming's and the complement cohorts). [d] $p$-value for estimates' difference testing between Deming (2009) replicated. and Complement cohort. For further details on models (1) to (5) specifications refer to Deming (2009).

**Table S16.** Comparison of Head Start Impacts for Deming's and Combined cohorts Unweighted/Weighted

|  | Replicated Deming | Weighted Deming[a] | Replicated Weighted Deming[b] | Combined cohorts | Weighted Combined cohorts |
|---|---|---|---|---|---|
| Adulthood summary index | .202*** (.072) | -- | .201*** (.075) | -.011 (.041) | -.038 (.046) |
| High school graduate | .087*** (.032) | .048 (.031) | .058* (.033) | .007 (.019) | -.008 (.022) |
| Idle | -.059* (.034) | -.055 (.037) | -.056 (.035) | .016 (.020) | .023 (.023) |
| Learning disability | -.059*** (.021) | -.042** (.018) | -.041** (.021) | .003 (.018) | -.004 (.013) |
| Poor health status | -.069*** (.027) | -.067** (.028) | -.065** (.027) | -.003 (.018) | .000 (.020) |
| Sample size | 1,251 | 1,251 | 1,251 | 3,738 | 3,738 |
| Head Start switchers | 581 | 581 | 581 | 1,928 | 1,928 |

*Notes:* The weighting scheme applied here follows that of Miller, Shenhav, & Grosz (MSG; 2019). [a] Deming's cohort weighted estimates, from MSG (2019). [b] Replication of MSG (2019). Standard error are in parenthesis. *** $p < .01$; ** $p < .05$; * $p < .1$.

**Table S17.** Head Start Impacts by Firstborn Status on Individual Outcomes, Across Cohorts

| | Grade retention | | Learning disability diagnosis | | Teen parenthood | | High school graduation | |
|---|---|---|---|---|---|---|---|---|
| | Deming | Combined | Deming | Combined | Deming | Combined | Deming | Combined |
| *Overall* | | | | | | | | |
| Head Start | -.092* | -.003 | -.064** | .004 | -.004 | .016 | .095** | .001 |
| | (.049) | (.026) | (.026) | (.013) | (.051) | (.025) | (.043) | (.003) |
| HS X First Born | .063 | -.006 | .013 | -.001 | -.031 | -.054 | -.023 | .003 |
| | (.078) | (.043) | (.041) | (.020) | (.083) | (.042) | (.069) | (.007) |
| Sample size | 1,169 | 3,512 | 1,247 | 3,707 | 1,251 | 3,738 | 1,251 | 3,738 |
| *< 4 years apart* | | | | | | | | |
| Head Start | -.047 | -.109 | .034 | .018 | .256 | .079 | -.044 | .032 |
| | (.226) | (.108) | (.084) | (.045) | (.166) | (.081) | (.135) | (.023) |
| HS X First Born | .541* | .213 | .177 | .052 | -.541** | -.130 | -.330 | -.009 |
| | (.300) | (.178) | (.161) | (.070) | (.233) | (.127) | (.267) | (.053) |
| Sample size | 184 | 442 | 198 | 463 | 199 | 466 | 199 | 466 |
| *< 5 years apart* | | | | | | | | |
| Head Start | -.195 | -.065 | -.022 | -.010 | .170* | .056 | .057 | .019 |
| | (.126) | (.080) | (.040) | (.039) | (.098) | (.058) | (.095) | (.017) |
| HS X First Born | .367* | .090 | .056 | .015 | -.320** | -.021 | -.150 | .016 |
| | (.212) | (.132) | (.104) | (.060) | (.162) | (.098) | (.178) | (.040) |
| Sample size | 274 | 674 | 298 | 716 | 300 | 722 | 300 | 722 |
| *< 6 years apart* | | | | | | | | |
| Head Start | -.040 | -.016 | -.070 | -.002 | .031 | .016 | -.035 | .010 |
| | (.097) | (.058) | (.048) | (.033) | (.083) | (.051) | (.080) | (.009) |
| HS X First Born | .065 | -.044 | .149* | .035 | -.030 | -.017 | -.047 | .008 |
| | (.154) | (.096) | (.088) | (.053) | (.134) | (.081) | (.113) | (.026) |
| Sample size | 429 | 992 | 457 | 1,056 | 459 | 1,062 | 459 | 1,062 |
| *< 7 years apart* | | | | | | | | |
| Head Start | -.050 | -.041 | -.041 | .012 | -.001 | .017 | -.014 | .005 |
| | (.080) | (.049) | (.040) | (.030) | (.071) | (.047) | (.066) | (.007) |
| HS X First Born | .085 | .012 | .052 | -.007 | .029 | -.012 | -.033 | .014 |
| | (.128) | (.080) | (.067) | (.044) | (.103) | (.072) | (.100) | (.025) |
| Sample size | 549 | 1,281 | 586 | 1,356 | 588 | 1,364 | 588 | 1,364 |

*Notes.* Model includes sibling fixed effects, indicators for Head Start, other pre-K, sex, age in 2014 and firstborn status. Pretreatment covariates and the interaction between Head Start and firstborn status are also included. This is the same model as our main analytic model with the addition of Head Start X Firstborn interaction to test for spillovers. 'No preschool' status is the counterfactual. Standard errors, in parenthesis, are clustered at the family level. *** $p < .01$; ** $p < .05$; * $p < .1$.

**Table S17.** Head Start Impacts by Firstborn Status on Individual Outcomes, Across Cohorts

| | Some college attended | | Idle | | Crime | | Poor health status | |
|---|---|---|---|---|---|---|---|---|
| | Deming | Combined | Deming | Combined | Deming | Combined | Deming | Combined |
| *Overall* | | | | | | | | |
| Head Start | .059 | .011 | -.068 | .023 | -.010 | .041 | -.068** | -.005 |
| | (.049) | (.027) | (.047) | (.025) | (.049) | (.026) | (.032) | (.023) |
| HS X First Born | -.001 | -.045 | -.009 | -.024 | .019 | -.097** | -.003 | .007 |
| | (.073) | (.044) | (.064) | (.040) | (.081) | (.045) | (.052) | (.036) |
| Sample size | 1,251 | 3,738 | 1,251 | 3,738 | 1,251 | 3,738 | 1,251 | 3,738 |
| *< 4 years apart* | | | | | | | | |
| Head Start | .432** | -.255* | .088 | -.012 | -.193 | -.181* | -.326*** | -.163* |
| | (.184) | (.140) | (.180) | (.080) | (.189) | (.099) | (.112) | (.093) |
| HS X First Born | -.444** | .095 | -.069 | .140 | .129 | .266 | .140 | .171 |
| | (.220) | (.204) | (.216) | (.113) | (.317) | (.168) | (.121) | (.121) |
| Sample size | 199 | 466 | 199 | 466 | 199 | 466 | 199 | 466 |
| *< 5 years apart* | | | | | | | | |
| Head Start | .243** | -.209** | -.010 | .020 | -.253** | -.181** | -.214** | -.047 |
| | (.105) | (.088) | (.095) | (.066) | (.113) | (.074) | (.083) | (.073) |
| HS X First Born | -.155 | .138 | .116 | .084 | .159 | .091 | .057 | .036 |
| | (.156) | (.128) | (.123) | (.109) | (.195) | (.130) | (.146) | (.101) |
| Sample size | 300 | 722 | 300 | 722 | 300 | 722 | 300 | 722 |
| *< 6 years apart* | | | | | | | | |
| Head Start | -.016 | -.117* | .005 | .035 | -.058 | -.008 | -.123** | -.024 |
| | (.089) | (.064) | (.070) | (.055) | (.074) | (.060) | (.061) | (.051) |
| HS X First Born | .084 | .119 | .074 | .059 | -.006 | -.130 | -.008 | .023 |
| | (.146) | (.099) | (.092) | (.097) | (.127) | (.102) | (.085) | (.075) |
| Sample size | 459 | 1,062 | 459 | 1,062 | 459 | 1,062 | 459 | 1,062 |
| *< 7 years apart* | | | | | | | | |
| Head Start | .023 | -.061 | -.063 | .026 | .001 | .028 | -.137** | -.058 |
| | (.080) | (.051) | (.071) | (.046) | (.072) | (.050) | (.054) | (.044) |
| HS X First Born | .003 | .036 | .076 | .054 | .028 | -.128 | .090 | .075 |
| | (.122) | (.081) | (.089) | (.078) | (.110) | (.083) | (.074) | (.066) |
| Sample size | 588 | 1,364 | 588 | 1,364 | 588 | 1,364 | 588 | 1,364 |

*Notes*. Model includes sibling fixed effects, indicators for Head Start, other pre-K, sex, age in 2014 and firstborn status. Pretreatment covariates and the interaction between Head Start and firstborn status are also included. This is the same model as our main analytic model with the addition of Head Start X Firstborn interaction to test for spillovers. 'No preschool' status is the counterfactual. Standard errors, in parenthesis, are clustered at the family level. *** $p < .01$; ** $p < .05$; * $p < .1$.

**Table S18.** Head Start Impacts by Firstborn Status on Indexes, Across Cohorts

| | Adulthood index | | Nontest index | |
|---|---|---|---|---|
| | Deming | Combined | Deming | Combined |
| *Overall* | | | | |
| Head Start | .240** | -.047 | .326*** | -.002 |
| | (.098) | (.053) | (.104) | (.058) |
| HS X First Born | .001 | .118 | -.156 | -.003 |
| | (.154) | (.096) | (.167) | (.097) |
| Sample size | 1,251 | 3,738 | 1,251 | 3,734 |
| *< 4 years apart* | | | | |
| Head Start | .452 | .182 | .279 | .100 |
| | (.390) | (.227) | (.411) | (.250) |
| HS X First Born | -.321 | -.372 | -1.773*** | -.574 |
| | (.552) | (.378) | (.604) | (.381) |
| Sample size | 199 | 466 | 199 | 466 |
| *< 5 years apart* | | | | |
| Head Start | .569** | .042 | .292 | .092 |
| | (.234) | (.170) | (.215) | (.186) |
| HS X First Born | -.256 | -.022 | -.759* | -.247 |
| | (.369) | (.296) | (.431) | (.292) |
| Sample size | 300 | 722 | 300 | 721 |
| *< 6 years apart* | | | | |
| Head Start | .114 | -.055 | .326 | .017 |
| | (.189) | (.133) | (.233) | (.141) |
| HS X First Born | -.012 | .151 | -.577* | -.052 |
| | (.265) | (.236) | (.347) | (.232) |
| Sample size | 459 | 1,062 | 459 | 1,061 |
| *< 7 years apart* | | | | |
| Head Start | .207 | -.020 | .226 | .007 |
| | (.169) | (.108) | (.193) | (.126) |
| HS X First Born | -.135 | .062 | -.284 | .001 |
| | (.153) | (.201) | (.278) | (.191) |
| Sample size | 588 | 1,364 | 588 | 1,363 |

*Notes.* Adulthood index (in standard deviation) = composite of 6 indicators for high school graduation; college attendance; teen-age parenthood; 'idle'; involvement with the justice system; and poor health status. Nontest index (in standard deviation) = composite of 2 indicators for learning disability; and grade repetition. Model includes sibling fixed effects, indicators for Head Start, other pre-K, sex, age in 2014 and firstborn status. Pretreatment covariates and the interaction between Head Start and firstborn status are also included. This is the same model as our main analytic model with the addition of Head Start X Firstborn interaction to test for spillovers. 'No preschool' status is the counterfactual. Standard errors, in parenthesis, are clustered at the family level. *** $p < .01$; ** $p < .05$; * $p < .1$.

**Table S19.** Head Start (HS) by cohort effect on adulthood summary index (measured in 2014), controlling for Head Start by covariate interaction(s)

|  | (HS x Cohort) main effect | HS x Covariate effect |
|---|---|---|
|  | -.263 |  |
|  | (.093) |  |
| *Head Start* x *Covariate:* |  |  |
| Pre-treatment index | -.255 | -.016 |
|  | (.098) | (.155) |
| Family human capital index | -.244 | -.057 |
|  | (.096) | (.057) |
| Gender | -.258 | -.123 |
|  | (.093) | (.098) |
| Black | -.262 | .040 |
|  | (.093) | (.092) |
| Low maternal AFQT | -.236 | .210 |
|  | (.093) | (.104) |
| Mother's age at child's birth | -.228 | -.005 |
|  | (.137) | (.014) |
| Mother's age at first child | -.272 | .006 |
|  | (.107) | (.013) |
| Mother's age at median child | -.275 | .004 |
|  | (.118) | (.014) |
| Mother's age at last child | -.316 | .014 |
|  | (.104) | (.009) |
| Age at measurement | -.232 | .003 |
|  | (.158) | (.015) |
| Family size: |  |  |
| 2 children | -.250 | -.110 |
|  | (.095) | (.109) |
| 3 children | -.259 | .024 |
|  | (.093) | (.096) |
| 4 children | -.261 | -.045 |
|  | (.093) | (.114) |
| 5+ children | -.252 | .142 |
|  | (.094) | (.112) |
| Adolescent during the Great Recession | -.276 | .024 |
|  | (.110) | (.123) |

*Notes.* Family clustered standard error are in parenthesis. Base model corresponds to Equation (1) of the text. Thus, sibling fixed-effects and pretreatment covariates are included throughout. Pre-treatment index: standardized sum of all standardized pre-treatment covariates. Family human capital: standardized sum of standardized maternal and grandparent's years of completed schooling; maternal AFQT score, family permanent income (ln), and CNLSY H.O.M.E. scale score. Low maternal AFQT standardized score (mean of 0; *SD* of 1): ≤-1. Mother's age at child's birth is a within-family variable. Mother's age at first, median, last child are between-families variables. Sample size = 3,141.

**Table S20.** Head Start Sibling Fixed Effect Impacts over 3 Birth Cohorts Grouping Samples

|  | Cohort 1 (born before 1983) | Cohort 2 (born btw 1983 & 1987) | Cohort 3 (born after 1987) |
| --- | --- | --- | --- |
| Adulthood summary index | 0.09 | -0.004 | -0.25* |
|  | (0.09) | (0.12) | (0.14) |
| $R^2$ | 0.73 | 0.81 | 0.77 |
| Sample size | 624 | 372 | 646 |
| Cognitive tests index | 0.01 | -0.15 | -0.09 |
|  | (0.08) | (0.11) | (0.10) |
| $R^2$ | 0.79 | 0.89 | 0.90 |
| Sample size | 619 | 363 | 620 |
| Nontests index | 0.29** | -0.10 | -0.06 |
|  | (0.10) | (0.15) | (0.15) |
| $R^2$ | 0.64 | 0.76 | 0.70 |
| Sample size | 624 | 372 | 645 |

*Notes*: Adulthood summary index (standardized) is a composite of 6 indicators: high school graduation; college attendance; teen-age parenthood; either working or attending school; involvement with the justice system; and poor health status. Cognitive test index (standardized) is a composite of the average of 3 standardized test scores: PPVT, PIAT Math, and PIAT Reading Recognition. Nontests index (standardized) comprises two indicators: grade retention and learning disability status. Impacts are expressed in standard deviation units. Standard errors are in parenthesis. ** $p < .05$; * $p < .1$. The counterfactual was a no preschool attendance.